\documentclass[11pt]{article}

\usepackage{amsmath,amsthm,amssymb}
\usepackage{fullpage}
\usepackage{authblk}

\usepackage[usenames,dvipsnames,svgnames,table]{xcolor}
\usepackage{graphicx}
\usepackage{caption}
\usepackage{subcaption}

\usepackage{tikz}
\usetikzlibrary{arrows}
\usetikzlibrary{positioning, fit}
\usetikzlibrary{shapes,snakes}
\usetikzlibrary{decorations.pathmorphing}
\usetikzlibrary{calc,through,intersections}
\tikzset{snake it/.style={decorate, decoration=snake}}

\title{Optimal diameter computation within bounded clique-width graphs}
\author[1,2]{Guillaume Ducoffe}
\affil[1]{\small National Institute for Research and Development in Informatics, Romania}
\affil[2]{\small University of Bucharest, Romania}
\date{}

\newtheorem{lemma}{Lemma}
\newtheorem{theorem}{Theorem}
\newtheorem{corollary}{Corollary}

\newtheorem{property}{Property}
\newtheorem{claim}{Claim}

\newtheorem{problem}{\bf Problem}
\newsavebox{\mybox}

\theoremstyle{definition}
\newtheorem{definition}{Definition}

\newcommand{\qedclaim}{\hfill $\diamond$ \medskip}
\newenvironment{proofclaim}{\noindent{\it Proof.}}{\qedclaim}

\begin{document}

\maketitle

\begin{abstract}
Coudert et al. ({\it SODA'18}) proved that under the Strong Exponential-Time Hypothesis, for any $\epsilon >0$, there is no ${\cal O}(2^{o(k)}n^{2-\epsilon})$-time algorithm for computing the diameter within the $n$-vertex cubic graphs of clique-width at most $k$. We present an algorithm which given an $n$-vertex $m$-edge graph $G$ and a $k$-expression, computes all the eccentricities in ${\cal O}(2^{{\cal O}(k)}(n+m)^{1+o(1)})$ time, thus matching their conditional lower bound. It can be modified in order to compute the Wiener index and the median set of $G$ within the same amount of time. On our way, we get a distance-labeling scheme for $n$-vertex $m$-edge graphs of clique-width at most $k$, using ${\cal O}(k\log^2{n})$ bits per vertex and constructible in ${\cal O}(k(n+m)\log{n})$ time from a given $k$-expression. Doing so, we match the label size obtained by Courcelle and Vanicat ({\it DAM 2016}), while we considerably improve the dependency on $k$ in their scheme. As a corollary, we get an ${\cal O}(kn^2\log{n})$-time algorithm for computing All-Pairs Shortest-Paths on $n$-vertex graphs of clique-width at most $k$. This partially answers an open question of Kratsch and Nelles ({\it STACS'20}).
\end{abstract}

\section{Introduction}\label{sec:intro}

For any undefined graph terminology, see~\cite{BoM08,Die10}.
Unless stated otherwise, all graphs considered in this work are simple, connected and unweighted (although we sometimes need to go beyond these assumptions in our proofs).
We here consider {\em clique-width}, that is one of the most studied parameters in Graph Theory, superseded only by the treewidth. Roughly, clique-width is a measure of the closeness of a graph to a cograph ({\it a.k.a.}, $P_4$-free graph). We postpone its formal definition until Sec.~\ref{sec:cut-tree}. The clique-width was shown to be bounded on many important subclasses of perfect graphs~\cite{BDHP16,BDHP17,DK16b,GoR00,LoR04}, and beyond~\cite{BLM04,BDLM05,BLM05,BELL06,BKM06,DK16,MaR99,SuT07,Van04}. For instance, distance-hereditary graphs, and so, trees, have clique-width at most three~\cite{GoR00}. Every graph of bounded treewidth also has bounded clique-width, but the converse is not true~\cite{CoR05}. Indeed, unlike for treewidth, there are dense graphs of bounded clique-width ({\it e.g.}, the complete graphs). This generality comes at some cost: whereas the celebrated Courcelle's theorem asserts that any problem expressible in MSO$_2$ logic can be solved in FPT linear time on bounded treewidth graphs~\cite{Cou90}, the same is true for bounded clique-width graphs only for the problems expressible in the more restricted MSO$_1$ logic~\cite{CMR00}. Fomin et al. showed this to be unavoidable, in the sense that there are problems expressible in MSO$_2$ logic that are $W[1]$-hard in the clique-width~\cite{FGLS10,FGLS14,FGLS+18}. We refer to~\cite{EGW01} for other algorithmic applications of clique-width in parameterized complexity.

\smallskip
Our focus is about the so-called ``FPT in P'' program. Here the goal is, for some problem solvable in ${\cal O}(m^{q+o(1)})$ time on arbitrary $m$-edge graphs, to design an ${\cal O}(f(k)m^{p+o(1)})$-time algorithm, for some $p < q$, within the class of graphs where some fixed parameter is at most $k$ (one usually seeks for $p=1$ and $f(k) = k^{{\cal O}(1)}$). The idea of using tools and methods from parameterized complexity in order to solve faster certain polynomial-time solvable problems has been here and there in the literature for a while ({\it e.g.}, see~\cite{HKNR98}). Nevertheless it was only recently that such idea was better formalized~\cite{GMN17}, in part motivated by some surprising results obtained for treewidth~\cite{AVW16}. Indeed, on the positive side, the treewidth does help in solving faster many important problems in P for graphs and matrices, that is, in $\tilde{\cal O}(k^{{\cal O}(1)}n)$ time on graphs and matrices of treewidth at most $k$~\cite{FLPS+17,IOO18}. But for other such problems, any truly subquadratic-time parameterized algorithm requires {\em exponential} dependency on the treewidth. In particular, recall that the distance $d_G(u,v)$ between two vertices $u$ and $v$ is equal to the least number of edges on a $uv$-path; the {\em diameter} of $G$ is defined as $diam(G) = \max_{u,v \in V} d_G(u,v)$. Abboud et al. proved that under the Strong Exponential-Time Hypothesis (SETH), for any $\epsilon > 0$, there is no ${\cal O}(2^{o(k)}n^{2-\epsilon})$-time algorithm for computing the diameter of $n$-vertex graphs of treewidth at most $k$~\cite{AVW16}. An algorithm in ${\cal O}(2^{{\cal O}(k)}n^{1+o(1)})$ time for this problem, thus matching the lower bound of Abboud et al., was proved recently in~\cite{BHM20} by using the orthogonal range query framework of Cabello and Knauer~\cite{CaK09}. For other aplications of this orthogonal range query framework to graph problems, see~\cite{Duc19,DHV19}. 

Insofar, clique-width has received less attention than treewidth in the nascent field of FPT in P. There is at least one good reason for that: unlike for treewidth, the parameterized complexity of clique-width is a wide open problem~\cite{CHLB+00}. Still, on many subclasses of bounded clique-width graphs, there exist linear-time algorithms in order to compute a so called ``$k$-expression'', for some $k = {\cal O}(1)$, with the latter certifying the clique-width of the graph to be at most $k$~\cite{GoR00,MaR99}. Therefore, the study of graph problems in P parameterized by clique-width may be regarded as a unifying framework for all such subclasses. In this respect, Coudert et al. obtained $\tilde{\cal O}(k^{{\cal O}(1)}(n+m))$-time algorithms for triangle and cycle problems on $n$-vertex $m$-edge graphs of clique-width at most $k$~\cite{CDP19}. However, they also observed that assuming SETH, even on $n$-vertex cubic graphs of clique-width at most $k$, for any $\epsilon > 0$, there is no ${\cal O}(2^{o(k)}n^{2-\epsilon})$-time algorithm for computing the diameter. Unlike for treewidth, it was open until this paper whether there does exist a parameterized quasi-linear-time algorithm for this problem on bounded clique-width graphs that matches their conditional lower bound. Their work has been continued in~\cite{DucoffeP18,DucoffeP18b,KratschN18} and especially in~\cite{kratsch_et_al:LIPIcs:2020:11899}, where the authors obtained an ${\cal O}((kn)^2)$-time algorithm for All-Pairs Shortest Paths (APSP) on $n$-vertex graphs of clique-width at most $k$. 

\paragraph{Results.} We provide several new insights on the fine-grained complexity of polynomial-time solvable distance problems within bounded clique-width graph classes.
\begin{enumerate}
\item Our main contribution is an ${\cal O}(2^{{\cal O}(k)}(n+m)^{1+o(1)})$-time algorithm for computing the diameter within the $n$-vertex $m$-edge graphs of clique-width at most $k$ (Theorem~\ref{thm:ecc}). To the best of our knowledge, it is the very first algorithm to match the conditional lower bound of Coudert et al. Furthermore, the same as for treewidth, we can easily modify our algorithm in order to compute other important distance invariants such as the Wiener index and the median set (Theorem~\ref{thm:td}), of which we recall their definition in Sec.~\ref{sec:diam}.
\item By using similar techniques, we obtain a new {\em distance labeling scheme} for bounded clique-width graphs classes which outperforms the state of the art~\cite{CourcelleV03}\footnote{In all fairness, the labeling scheme of Courcelle and Vanicat can be applied to many more problems than just the computation of the distances in the graph.}. See our Theorem~\ref{thm:dls-new} for details. In doing so, we immediately get an ${\cal O}(kn^2\log{n})$-time algorithm in order to solve All-Pairs Shortest-Paths within $n$-vertex graphs of clique-width at most $k$ (Corollary~\ref{cor:apsp}). This almost completely solves an open problem from Kratsch and Nelles~\cite{kratsch_et_al:LIPIcs:2020:11899} who asked whether there exists an ${\cal O}(kn^2)$-time algorithm for this problem.
\end{enumerate}

\paragraph{Overview of our techniques.}
Our high-level approach for computing the diameter is rather standard: we disconnect the input graph into balanced subgraphs, we use the framework of Cabello and Knauer in order to compute the maximum distance between vertices that are on different subgraphs, then we end up applying our algorithm recursively to each subgraph. However, instead of disconnecting the graph with small balanced separators -- as it was done for treewidth in~\cite{AVW16,BHM20} -- we rather use edge-cuts of small neighbourhood diversity. As a side contribution of this work, we show how to compute such balanced cuts in linear time from a given $k$-expression. This result seems to be already known~\cite{BJRS02,DraganY10}, but we were unable to find the reference. Here, an important step in our approach consists in first transforming a $k$-expression into a so called {\em partition tree}, a purely combinatorial object which has been used in~\cite{CHMPR15} in order to derive a new characterization of the clique-width. -- Interestingly, we may regard our distance-labeling scheme of Theorem~\ref{thm:dls-new} as a modified centroid decomposition of this partition tree. -- Then, we show in Sec.~\ref{sec:rq} that the orthogonal range query framework can be applied in order to compute the maximum distance between any two vertices that are separated by a cut of small neighbourhood diversity. 

From this point on, a more subtle complication occurs due to the need to ``repair'' the distances in the subgraphs on which we want to recurse so that they coincide with the distances in the original graph. For bounded treewidth graphs, previous works have remedied to that issue by adding some {\em weighted edges} between the vertices in the small balanced separators. Unfortunately, we {\em cannot} solve the diameter problem within weighted graphs of bounded clique-width. In fact, as it was observed in~\cite{KratschN18}, most problems on bounded clique-width weighted graphs are as hard as on general weighted graphs. This is because we may regard any graph as a weighted clique, where each non-edge got replaced by an edge of sufficiently large weight. Our solution here consists in ensuring that all the weighted edges in the subgraphs considered can be partitioned in at most ${\cal O}(\log{n})$ clusters of only ${\cal O}(k^2)$ vertices; in particular, each cluster can have all its vertices put on the same side of an edge-cut, while preserving the property for the latter to be balanced {\em and} not increasing the neighbourhood diversity of the cut. Doing so, all edges of the cuts considered stay unweighted. 

Note that for the distance-labeling scheme of Theorem~\ref{thm:dls-new} we avoid dealing with weighted graphs, that is why we deem the proof of this result simpler than the ones of Theorems~\ref{thm:ecc} and~\ref{thm:td}, and we chose to present it first in the paper. However, doing so, we are left dealing with possibly unconnected graphs. While it is likely that we could process each connected component separately, we did not explore this possibility since it was leading to more complicated updates of the partition trees.   

\paragraph{Notations.} We now state the basic terminology used in the paper, in a sufficiently general way so that it can be applied to all types of graphs considered in our proofs. By a {\em weighted graph}, we mean a triple $G=(V,E,w)$ where $w : E \to \mathbb{N}$. If for every edge $e$ we have $w_e = 1$, then we call the graph {\em unweighted} and we simply write it $G=(V,E)$. The neighbour set of a vertex $v \in V$, resp. of a subset $S \subseteq V$, is defined as $N_G(v) = \{ u \in V \mid uv \in E \}$, resp. as $N_G(S) = \bigcup_{v \in S} N_G(v) \setminus S$. The {\em distance} $d_G(u,v)$ between $u,v \in V$ is equal to: $+\infty$ if $u$ and $v$ are on different connected components of $G$, and to the smallest weight of a $uv$-path in $G$ otherwise (resp., if $G$ is unweighted, to the least number of edges of a $uv$-path). We may also define the distance between a vertex $v \in V$ and a subset $S \subseteq V$ as $d_G(v,S) = d_G(S,v) = \min_{u \in S} d_G(u,v)$, and the distance between two subsets $S,S'$ as $d_G(S,S') = \min_{u \in S, v \in S'} d_G(u,v)$. Note that if $S = \emptyset$ then, $d_G(v,S) = d_G(S,S') = +\infty$ for any $v$ and $S'$. Finally, we recall that the diameter of $G$ is equal to $diam(G) = \max_{u,v \in V} d_G(u,v)$. We introduce additional terminology where it is needed in the paper.

\section{Clique-width and partition trees}\label{sec:cut-tree}

We start recalling the definition of clique-width, then we sketch the transition from a $k$-expression to a partition tree~\cite{CHMPR15}. We end up this section with a few nice properties of partition trees.

\paragraph{Clique-width expressions.} A $k$-labeled graph is a triple $G=(V,E,\ell)$ where $\ell : V \to \{1,2,\ldots,k\}$ is called a labeling function. A clique-width $k$-expression (for short, a $k$-expression) is an algebraic expression where the four allowed operations are:
\begin{itemize}
\item $i(v)$: we add a new isolated vertex with label $\ell(v) = i$;
\item $G_1 \oplus G_2$: we make the disjoint union of two $k$-labeled graphs;
\item $\eta(i,j)$: we add a join (complete bipartite subgraph) between all vertices with label $i$ and all vertices with label $j$;
\item $\rho(i,j)$: for all vertices $v$ s.t. $\ell(v) = i$, we set $\ell(v) = j$.
\end{itemize}
The generated graph is the one obtained from the $k$-expression by deleting all the labels. We say that a graph $G=(V,E)$ has clique-width at most $k$ if it is the graph generated by some $k$-expression. For instance, $1(a)2(b)\eta(1,2)\rho(1,3)1(c)\eta(1,2)\rho(2,3)2(d)\eta(1,2)$ is a $3$-expression generating the four-node path $P_4$ with nodes $a,b,c,d$. In particular, the clique-width of $P_4$ is at most three. This is in fact an equality, as the graphs of clique-width at most two are exactly the cographs~\cite{GoR00}. We denote by $cw(G)$ the clique-width of the graph $G$. The size of a $k$-expression is its number of operations. If the generated graph has order $n$ and $m$ edges, and there is no unnecessary operation $\rho(i,j)$ -- which we will assume to be the case throughout the remainder of this paper --, then the $k$-expression has size in ${\cal O}(n+m)$ ({\it e.g.}, see~\cite{Furer14}, where F\"urer proved this result for the more general notion of $k$-fusion-tree expression).

\begin{figure}[h!]
	\begin{center}
		\begin{subfigure}[b]{.46\textwidth}\centering
			\begin{tikzpicture}
				\node[label={$\eta(1,2)$}] at (0,0) {};
				\node[label={$\oplus$}] at (0,-1) {};
				\node[label={$2(d)$}] at (1,-2) {};
				\node[label={$\rho(2,3)$}] at (-1,-2) {};
				\node[label={$\eta(1,2)$}] at (-1,-3) {};
				\node[label={$\oplus$}] at (-1,-4) {};
				\node[label={$1(c)$}] at (0,-5) {};
				\node[label={$\rho(1,3)$}] at (-2,-5) {};
				\node[label={$\eta(1,2)$}] at (-2,-6) {};
				\node[label={$\oplus$}] at (-2,-7) {};
				\node[label={$1(a)$}] at (-3,-8) {};
				\node[label={$2(b)$}] at (-1,-8) {};

				\draw (0,0) -- (0,-.4);
				\draw (-1,-1.3) -- (0,-.8) -- (1,-1.3);
				\draw (-1,-1.8) -- (-1,-2.3);
				\draw (-1,-2.8) -- (-1,-3.3);
				\draw (-2,-4.3) -- (-1,-3.8) -- (0,-4.3);
				\draw (-2,-4.8) -- (-2,-5.3);
				\draw (-2,-5.8) -- (-2,-6.3);
				\draw (-3,-7.3) -- (-2,-6.8) -- (-1,-7.3);
			\end{tikzpicture}
			\caption{Syntactic tree for a $3$-expression of $P_4$.}
			\label{fig:syntactic-tree}
		\end{subfigure}\hfill
		\begin{subfigure}[b]{.46\textwidth}\centering
			\begin{tikzpicture}
				\node[label={$\{\{c\},\{d\},\{a,b\}\}$}] at (0,0) {};
				\node[label={$\{\{d\}\}$}] at (2,-2) {};
				\node[label={$\{\{c\},\{b\},\{a\}\}$}] at (-2,-2) {};
				\node[label={$\{\{a\},\{b\}\}$}] at (-3,-4) {};
				\node[label={$\{\{c\}\}$}] at (-1,-4) {};
				\node[label={$\{\{a\}\}$}] at (-4,-6) {};
				\node[label={$\{\{b\}\}$}] at (-2,-6) {};
				
				\draw (2,-1.3) -- (0,0) -- (-2,-1.3);
				\draw (-3,-3.3) -- (-2,-1.8) -- (-1,-3.3);
				\draw (-4,-5.3) -- (-3,-3.8) -- (-2,-5.3);
			\end{tikzpicture}
			\caption{A corresponding partition tree.}
			\label{fig:partition-tree}
		\end{subfigure}	
	\end{center}
\end{figure}

\paragraph{Partition tree.} It is useful to represent a $k$-expression as a syntactic tree. See Fig.~\ref{fig:syntactic-tree} for an illustration. By iteratively contracting the edges incident to non-branching nodes, we get a so-called partition tree, whose nodes are mapped to the collection of subsets of vertices with equal label in their rooted subtree. See Fig.~\ref{fig:partition-tree}. Formally, given a graph $G=(V,E)$, a partition tree is a pair $(T,f)$ where $T$ is a rooted tree whose all inner nodes have at least two children, such that:
\begin{itemize}
\item for every node $a \in V(T)$, $f(a)$ is a partition of some vertex-subset $A \subseteq V$;
\item for every vertex $v \in V$, there is a leaf node $a_v \in V(T)$ s.t. $f(a_v) = \{\{v\}\}$;
\item for every inner node $a \in V(T)$, let $b_1,b_2,\ldots,b_d$ be its children. If $f(a)$ is a partition of $A $, and in the same way for every $1 \leq i \leq d$, $f(b_i)$ is a partition of $B_i$, then the vertex-subsets $B_1,B_2,\ldots,B_d$ are pairwise disjoint and $A = \bigcup_{i=1}^k B_i$. Furthermore:
\begin{itemize}
\item For every $1 \leq i \leq d$, for every subset $X_i \in f(b_i)$, there is a $X \in f(a)$ s.t. $X_i \subseteq X$ (we say that $\bigcup_{i=1}^d f(b_i)$ refines $f(a)$);
\item Finally, for every $1 \leq i < j \leq d$, for every adjacent vertices $v_i \in B_i$ and $v_j \in B_j$, if $v_i \in X$ and $v_j \in Y$, for some $X,Y \in f(a)$, then we have $X \neq Y$ and $X \times Y \subseteq E$ (we say that the partition is compatible with the edge-incidence relation in the graph $G$). 
\end{itemize} 
\end{itemize}
The {\em width} of a partition tree is equal to $\max_{a \in V(T)} |f(a)|$. A graph has clique-width at most $k$ if and only if it admits a partition tree of width at most $k$~\cite{CHMPR15}. 

Note that if we naively store a partition tree $(T,f)$, then storing explicitly all the labels $f(a)$, for $a \in V(T)$, would require ${\cal O}(n^2)$ space. Instead, for every $a \in V(T)$, for every $X \in f(a)$, we may create a new vertex $(a,X)$; then if $b_i$ is a child of $a$, for every $X_i \in f(b_i)$ s.t. $X_i \subseteq X$, we add an arc between $(a,X)$ and $(b_i,X_i)$. This is called in~\cite{CHMPR15} the representation graph of $(T,f)$ and it only requires ${\cal O}(kn)$ space if the width is at most $k$.

\begin{lemma}[\cite{CHMPR15}]\label{lem:partition-tree}
There is an algorithm that transforms a $k$-expression of size $L$ into the representation graph of a width-$k$ partition tree in ${\cal O}(kL)$ time.
\end{lemma}

In particular, given a $k$-expression for an $n$-vertex $m$-edge graph $G$, we can construct the representation graph of a width-$k$ partition tree in ${\cal O}(k(n+m))$ time.

\paragraph{Relation with $k$-modules.}
For a graph $G=(V,E)$, a subset $M \subseteq V$ is a module if we have $N_G(u) \setminus M = N_G(v) \setminus M$ for every vertices $u,v \in M$. A $k$-module is some $M \subseteq V$ that can be partitioned into $k$ subsets, denoted $M_1,M_2,\ldots,M_k$, in such a way that for every $1 \leq i \leq k$, $M_i$ is a module in the subgraph $G[ (V\setminus M) \cup M_i ]$. Some relations between clique-width and $k$-modules were explored in~\cite{Rao08}. We make the following useful observation, whose proof is inspired by~\cite[Theorem $7$]{Rao08}.

\begin{lemma}\label{lem:k-mod}
The following two properties hold for every partition tree $(T,f)$ of a graph $G=(V,E)$:
\begin{enumerate}
\item For every node $a \in V(T)$, let $A = \bigcup f(a)$ be the vertex-subset of which $f(a)$ is a partition. Then, $A$ is a $|f(a)|$-module of $G$, with a corresponding partition of $A$ being $f(a)$.
\item Let $a_1,a_2,\ldots,a_p$ be children nodes of some $a' \in V(T)$ and, for every $1 \leq i \leq p$, let $A_i = \bigcup f(a_i)$ be the vertex-subset of which $f(a_i)$ is a partition. Then, $A = \bigcup_{i=1}^p A_i$ is a $|f(a')|$-module of $G$, with a corresponding partition of this subset being $\{ X' \cap A \mid X' \in f(a') \}$. 
\end{enumerate}
\end{lemma}

\begin{proof}
We prove these two above statements simultaneously, by induction on the depth of the nodes. For the base case, let us assume $a \in V(T)$ to be the root of $T$. In particular, $A = V$. In this situation, every $X \in f(a)$ is a trivial module of $G \setminus (V \setminus X) = G[X]$. Hence, $A = V$ is a $k$-module of $G$ with a corresponding partition being $f(a)$. Then, let $a_1,a_2,\ldots,a_p$ be children nodes of some $a' \in V(T)$, and let us assume by induction that $A' = \bigcup f(a')$ is a $|f(a')|$-module of $G$, with a corresponding partition being $f(a')$. Recall that $A = \bigcup_{i=1}^p A_i$ is the union of all the subsets partitioned by the $f(a_i)$'s. By the refinement property we have $A \subseteq A'$, and therefore $\Phi(A) = \{ X' \cap A \mid X' \in f(a') \}$ is a partition of $A$. Let us prove that $A$ is a $|\Phi(A)|$-module of $G$, with a corresponding partition being $\Phi(A)$ (Property $2$ of the lemma). Equivalently, we are left proving that for every $X \in \Phi(A)$, for every $u,v \in X$ we have $N_G(u) \setminus A = N_G(v) \setminus A$. For that, recall that there is a $X' \in f(a')$ s.t. $X \subseteq X'$. By our induction hypothesis, $X'$ is a module of $G \setminus (A' \setminus X')$. Therefore, $N_G(u) \setminus A' = N_G(v) \setminus A'$. In order to prove that $X$ is a module of $G \setminus (A \setminus X)$, it now suffices to prove that we have $N_G(u) \cap (A' \setminus A) = N_G(v) \cap (A' \setminus A)$. Let $w \in A' \setminus A$ be s.t. $uw \in E$. The refinement property implies the existence of some node $b \notin \{a_1,a_2,\ldots,a_p\}$ s.t. $b$ is another child of $a'$, $f(b)$ is a partition of some vertex-subset $B$ that is disjoint from $A$, and $w \in B$. Then, since $uw \in E$, the compatibility property implies the existence of some $Y' \in f(a')$ s.t. $Y' \neq X'$, $w \in Y'$ and $X' \times Y' \subseteq E$. In particular, every vertex of $X'$, and so, of $X$, is adjacent to $w$. This implies $N_G(u) \cap (A' \setminus A) = N_G(v) \cap (A' \setminus A)$. Finally, let us prove that for every child $a$ of $a'$, $A = \bigcup f(a)$ is also a $|f(a)|$-module of $G$, with a corresponding partition being $f(a)$ (Property $1$ of the lemma). By setting $p=1$, we first get that $A$ is a $|f(a')|$-module, with a corresponding partition being $\Phi(A) = \{ X' \cap A \mid X \in f(a') \}$. Then, we are done by the refinement property because every $X \in f(a)$ must be contained into some $X' \cap A \in \Phi(A)$.
\end{proof}

Finally, recall that a cut of a graph $G=(V,E)$ is a bipartition $(A,V \setminus A)$ of its vertex-set. The {\em neighbourhood diversity} of a cut is the least $k$ s.t. $A$ is a $k$-module of $G$. It follows from Lemma~\ref{lem:k-mod} that every node of a width-$k$ partition tree defines a cut of neighbourhood diversity at most $k$.

\section{Distance-labeling scheme}\label{sec:dls}

We start describing our distance oracle for bounded clique-width graph classes as it is a bit simpler to present than our results for the diameter problem.
For a graph $G=(V,E)$, a {\em distance-labeling scheme} consists in some encoding function $C_G : V \to \{0,1\}^*$ and some decoding function $D_G : \{0,1\}^* \times \{0,1\}^* \to \mathbb{N} \cup \{+\infty\}$ s.t. $d_G(u,v) = D_G(C_G(u),C_G(v))$ for every vertices $u$ and $v$. We are interested in minimizing the total pre-processing time in order to compute the labels $C_G(v)$, for all vertices $v$, and the query time in order to compute the distance given two labels. It is often the case that $D_G$ runs in time polynomial in the size of the labels. Then, the objective is to minimize the maximum bit size of the labels, {\it i.e.}, $\max_{v \in V} |C_G(v)|$. 

The following result is due to Courcelle and Vanicat:

\begin{theorem}[\cite{CourcelleV03}]\label{thm:dls-state-of-the-art}
The family of $n$-vertex bounded clique-width graphs enjoys an exact distance labeling scheme using labels of length ${\cal O}(\log^2{n})$ bits. Moreover, the distance can be computed in ${\cal O}(\log^2{n})$ time.
\end{theorem}

The hidden dependency in the clique-width is a stack of exponentials~\cite{gavoille2003distance}. We improve the latter while keeping optimal bit size and improved query time, namely:

\begin{theorem}\label{thm:dls-new}
The family of $n$-vertex $m$-edge graphs of clique-width at most $k$ enjoys an exact distance labeling scheme using labels of length ${\cal O}(k\log^2{n})$ bits. Moreover, all the labels can be pre-computed in ${\cal O}(k(n+m)\log{n})$ time if a $k$-expression is given, and the distance can be computed in ${\cal O}(k\log{n})$ time.
\end{theorem} 

Recall that $d_G(v,S) = d_G(S,v) = \min_{u \in S} d_G(u,v)$. In particular, $d_G(v,S) = +\infty$ if $S$ is empty. 
We will need the following result:

\begin{lemma}\label{lem:dist-formula}
Let $G=(V,E)$ be a graph (possibly unconnected) and let $(A,V \setminus A)$ be a cut of neighbourhood diversity at most $k$. Furthermore, let $A_1,A_2,\ldots,A_k$ be a partition of $A$ s.t. for every $1 \leq i \leq k$, $A_i$ is a module of $G \setminus (A \setminus A_i)$. For $1 \leq i \leq k$, let $B_i = N_G(A_i) \setminus A$. The following hold for every $u,v \in V$:
\begin{itemize}
\item if $u \in A, \ v \notin A$ then $d_G(u,v) = \min\{ d_G(u,A_i) + 1 + d_G(B_i,v) \mid 1 \leq i \leq k \}$;
\item if $u,v \in A$ then $d_G(u,v) = \min \{d_{G[A]}(u,v)\} \cup \{ d_G(u,A_i) + 1 + d_G(B_i,v) \mid 1 \leq i \leq k \}$;
\item if $u,v \notin A$ then $d_G(u,v) = \min \{d_{G[V \setminus A]}(u,v)\} \cup \{ d_G(u,A_i) + 1 + d_G(B_i,v) \mid 1 \leq i \leq k \}$.
\end{itemize}
\end{lemma}
\begin{proof}
We may assume $u$ and $v$ to be in a same connected component of $G$. Indeed if it is not the case then we claim that, for any $1 \leq i \leq k$, we have $d_G(u,A_i) = +\infty$ or $d_G(B_i,v) = +\infty$; in particular, the lemma holds true in this special case. In order to prove this claim, there are two simple cases to consider: either $B_i = \emptyset$, and then $d_G(B_i,v) = +\infty$, or $B_i \neq \emptyset$. In the latter case, $A_i \cup B_i$ is connected, and therefore we must have $d_G(u,A_i) = +\infty$ or $d_G(B_i,v) = +\infty$. From now on, we implicitly assume the existence of a $uv$-path.
Then, for {\em any} $u$ and $v$, for every $1 \leq i \leq k$ s.t. $B_i \neq \emptyset$, since there is a complete join between $A_i$ and $B_i$ there always exists a $uv$-path of length $d_G(u,A_i) + 1 + d_G(B_i,v)$ (recall that if $B_i = \emptyset$, then we have $d_G(u,A_i) + 1 + d_G(B_i,v) = d_G(B_i,v) = +\infty$). In particular, $d_G(u,v) \leq \min\{ d_G(u,A_i) + 1 + d_G(B_i,v) \mid 1 \leq i \leq k \}$. Then, we consider all three cases of the lemma. If $u \in A, \ v \notin A$ then on any shortest $uv$-path, there must be some edge $u'v'$ s.t. $u' \in A, \ v' \notin A$. In particular, $u' \in A_i$ for some $1 \leq i \leq k$, and then $v' \in B_i$. We get $d_G(u,v) \geq d_G(u,A_i) + 1 + d_G(B_i,v)$. As a result, $d_G(u,v) = \min\{ d_G(u,A_i) + 1 + d_G(B_i,v) \mid 1 \leq i \leq k \}$. If $u,v \in A$ then, either there exists a shortest $uv$-path which is fully contained into $A$, that implies $d_G(u,v) = d_{G[A]}(u,v)$, or {\em every} shortest $uv$-path must intersect $V \setminus A$. In the latter sub-case, we fix a shortest $uv$-path and we scan it from $u$ until we find an edge $u'v'$ s.t. $u' \in A, v' \notin A$. We deduce as before that we have in this sub-case $d_G(u,v) = \min\{ d_G(u,A_i) + 1 + d_G(B_i,v) \mid 1 \leq i \leq k \}$. In the same way, if $u,v \notin A$ then either there exists a shortest $uv$-path which is fully out of $A$, that implies $d_G(u,v) = d_{G[V \setminus A]}(u,v)$, or {\em every} shortest $uv$-path must intersect $A$. In the latter sub-case, we fix a shortest $uv$-path and we scan it from $v$ until we find an edge $v'u'$ s.t. $u' \in A, v' \notin A$.
\end{proof}

Our scheme for bounded clique-width graphs mimics one very well-known for trees which is based on the centroid decomposition~\cite{GPPR0404}. Specifically, let $w : V(T) \to \mathbb{N}$ assign non-negative weights to the nodes of some tree $T$. A $w$-centroid is a node $c$ s.t. every subtree of $T \setminus \{c\}$ has weight at most $w(T)/2$. Such node always exists and a centroid can be computed in linear time by using a standard dynamic programming approach~\cite{Gol71}. We will also need the following easy lemma:

\begin{lemma}\label{lem:bipartition-centroid}
If $c$ is a $w$-centroid of a tree $T$, then the components of $T \setminus \{c\}$ can be partitioned in linear-time in two forest $F_1,F_2$ s.t. $\max\{w(F_1),w(F_2)\} \leq 2w(T)/3$.
\end{lemma}

\begin{proof}
The result is trivial if $T \setminus \{c\}$ is connected ({\it i.e.}, we set $F_1 = T \setminus \{c\}, \ F_2 = \emptyset$). In the same way, if $w(c) > w(T)/3$ then the result holds for any bipartition of the components of $T \setminus \{c\}$. From now on we assume that we did not fall in one of those two pathological cases. Let $T_1,T_2,\ldots,T_d$ be the subtrees of $T \setminus \{c\}$. We define $i_0$ as the least index $i$ s.t. $\sum_{j=1}^i w(T_j) > 2w(T)/3$. Note that $i_0 > 1$ since we assume $c$ to be a $w$-centroid. Then, there are two cases. 
\begin{itemize}
\item If $\sum_{j=i_0}^d w(T_j) \leq 2w(T)/3$ then we are done by setting $F_1 = \bigcup_{j < i_0} T_j, \ F_2 = \bigcup_{j \geq i_0} T_j$.
\item Otherwise, we get $w(T) + w(T_{i_0}) \geq \sum_{j \leq i_0} w(T_j) + \sum_{j \geq i_0} w(T_j) > 4w(T)/3$, and so, $w(T_{i_0}) > w(T)/3$. We set $F_1 = T_{i_0}, \ F_2 = \bigcup_{j \neq i_0} T_j$.  
\end{itemize}
In both cases, we get the desired partition in two forests of respective weights at most $2w(T)/3$.
\end{proof}

We are now ready to prove the main result of this section:

\begin{proof}[Proof of Theorem~\ref{thm:dls-new}]
We fix some width-$k$ partition tree $(T,f)$, that takes ${\cal O}(k(n+m))$ time by using Lemma~\ref{lem:partition-tree}. Let $w : V(T) \to \{0,1\}$ be s.t. $w(a) = 1$ if and only if $a$ is a leaf. Observe that $w(T) = n$ since there is a one-to-one mapping between the vertices in $V$ and the leaves of $T$. In order to construct the labels $C_G(v)$, for all $v \in V$ (encoding function), we next define a recursive procedure onto the weighted partition tree.

\medskip
In what follows, let us assume $n > 1$ (otherwise, there is nothing to be done).
We compute in ${\cal O}(|V(T)|)$ time, and so in ${\cal O}(n)$ time, a $w$-centroid $c$. 
Note that if $n=2$, then $T$ is composed of a root and of two leaves; then, a good choice for the $w$-centroid $c$ is to take the root. In particular, we may assume $c$ to be an internal node. Otherwise, $n \geq 3$, and so, since $w(T) = n$, we {\em must} have that $c$ is an internal node. Then, let $a_1,a_2,\ldots,a_d$ be the children of $c$. We denote $C$ (resp. $A_i$) the subset of vertices of which $f(c)$ (resp., $f(a_i)$) is a partition. Furthermore, let $T_c$ (resp., let $T_{a_i}$) be the subtree rooted at $c$ (resp., at $a_i$). By Lemma~\ref{lem:bipartition-centroid} we can bipartition the trees $T \setminus T_c, \ T_{a_1},T_{a_2},\ldots,T_{a_d}$ into two forests $F_1,F_2$ of respective total weights $\leq 2n/3$. In particular, since $c$ is internal, and so $w(c) = 0$, both forests are non-empty. Up to re-ordering the children nodes of $c$, we may assume one of those forests, say $F_1$, to be equal to $\bigcup_{j=1}^p T_{a_j}$, for some $p \leq d$. Doing so, we define the cut $\left(\bigcup_{j=1}^p A_j, V \setminus \left(\bigcup_{j=1}^p A_j\right)\right)$, whose two sides can be determined in ${\cal O}(n)$ time by traversing the disjoint subtrees $T_{a_1},T_{a_2},\ldots,T_{a_p}$. For short, we name $A := \bigcup_{j=1}^p A_j$.

By Lemma~\ref{lem:k-mod}, $A$ is a $k$-module of $G$, with a corresponding partition being $\Phi(A) = \{ X \cap A \mid X \in f(c) \}$ (or $f(a_1)$ if $p=1$). Note that such a partition can be readily derived in ${\cal O}(n)$ time from either $f(c)$ or $f(a_1)$. In turn, being given the representation graph of $(T,f)$, we can compute $f(c)$ and $f(a_1)$ in ${\cal O}(kn)$ time by traversing the subtrees rooted at nodes $c$ and $a_1$.
Let $X_1, X_2, \ldots, X_k$ be a partition of $A$ s.t., for every $1 \leq i \leq k$, $X_i$ is a module of $G \setminus (A \setminus X_i)$. Furthermore, for every $1 \leq i \leq k$, let $Y_i := N_G(X_i) \setminus A$ (neighbour sets in $V \setminus A$). Since the subsets $X_i$ are pairwise disjoint we can compute $Y_1,Y_2,\ldots,Y_k$ in total ${\cal O}(m)$ time. Finally, for every $1 \leq i \leq k$, for every $v \in V$, we compute $d_G(v,X_i)$ and $d_G(v,Y_i)$. It takes ${\cal O}(m+n)$ time per subset, using a modified BFS, and so total time in ${\cal O}(k(m+n))$. We end up applying recursively the same procedure as above on the disjoint (possibly unconnected) subgraphs $G[A]$ and $G[V \setminus A]$. For that, we need to build a partition tree for each subgraph. 
\begin{itemize}
\item For $G[A]$, we take $T_A = T_{a_1}$ if $p=1$, otherwise we take $T_A = T_c \setminus (\bigcup_{j > p}T_j)$. Then, for every $b \in V(T_A)$, we set $f_A(b) = \{ X \cap A \mid X \in f(b) \}$. Observe that if $b \neq c$ then $f_A(b) = f(b)$. Hence, the representation graph of $(T_A,f_A)$ can be computed from the representation graph of $(T,f)$ in ${\cal O}(kn)$ time.
\item For $G[V \setminus A]$, a natural choice would be to take the subtree $T_{V \setminus A} = T \setminus (\bigcup_{j=1}^p T_{a_j})$. Then, for every $b \in V(T_{V\setminus A})$, we set $f_{V \setminus A}(b) = \{ X \setminus A \mid X \in f(b) \}$. Again, we observe that the representation graph of $(T_{V \setminus A}, f_{V \setminus A})$ can be computed from the representation graph of $(T,f)$ in ${\cal O}(kn)$ time. However, doing so, we may not respect all properties of a partition tree. Specifically, if $d=p$ then $c$ has become a leaf-node and it must be removed. But then, its father node $c'$ may have only one child $b$ left. If that is the case, then either $c'$ is the root of $T$ and then we choose $T_{V \setminus A} = T_b$, or we choose the father node of $c'$ as the new father node of $b$, removing on our way the node $c'$. Note that we do {\em not} modify $f_{V \setminus A}(b)$ during this procedure. Finally, if $d = p+1$ then $c$ only has one child $a_d$ left. We proceed similarly as in the previous case. That is, either $c$ was the root of $T$ and then we set $T_{V \setminus A} = T_{a_d}$, or we choose the father node of $c$ as the new father node of $a_d$, removing on our way the node $c$. Note that we do {\em not} modify the partition $f_{V \setminus A}(a_d)$ during this procedure. 
\end{itemize}

\medskip
The above procedure recursively defines a so called $w$-centroid decomposition $T^{(w)}$. The latter is a binary rooted tree, whose root is labeled by the cut $(A,V \setminus A)$. Its left and right subtrees are $w$-centroid decompositions of $G[A]$ and $G[V \setminus A]$ respectively. Note that by construction, the depth of $T^{(w)}$ is in ${\cal O}(\log{n})$. Furthermore, there is a one-to-one mapping between the leaves of $T^{(w)}$ and the vertices of $G$. For every vertex $v \in V$, its label $C_G(v)$ contains the $2k$ distances computed for each cut on its path until the root of $T^{(w)}$. -- Infinite distances may be encoded as some special character. -- Here, we stress that all these distances are computed in some induced subgraphs of $G$, and not in $G$ itself (unless it is for the first cut, at the root). Since the depth of $T^{(w)}$ is in ${\cal O}(\log{n})$, each $C_G(v)$ stores ${\cal O}(k\log{n})$ distances, and so it has a bit size in ${\cal O}(k\log^2{n})$. Furthermore, as $G[A]$ and $G[V \setminus A]$ are disjoint, every recursive stage of the procedure takes ${\cal O}(k(n+m))$ time. Hence, the total pre-processing time in order to compute $C_G(v)$, for all $v \in V$, is in ${\cal O}(k(n+m)\log{n})$.

We are left describing $D_G$ (decoding). Let $u,v \in V$ be arbitrary. Their least common ancestor in $T^{(w)}$ corresponds to some cut $(A^j,A^{j-1}\setminus A^j)$ s.t. $u \in A^j, \ v \in A^{j-1} \setminus A^j$. Consider {\em all} the cuts on the path between their least common ancestor and the root of $T^{(w)}$. We call the latter $(A^0,V\setminus A^0), (A^1,A^0 \setminus A^1), \ldots, (A^j,A^{j-1}\setminus A_j)$. Since up to reverting their two sides, all these cuts have neighbourhood diversity at most $k$, then we may apply Lemma~\ref{lem:dist-formula} $j$ times in order to compute $d_G(u,v)$ ({\it i.e.}, in $G, G[A^0], G[A^1], \ldots, G[A^{j-1}]$). Note that $j = {\cal O}(\log{n})$. Finally, since for each cut considered, the $2k$ distances that are required in order to apply this lemma are stored in $C_G(u)$ and $C_G(v)$, it takes ${\cal O}(k)$ time per cut, and so, the final query time is in ${\cal O}(k\log{n})$.
\end{proof}

Recall that All-Pairs Shortest-Paths in an $n$-vertex graph of clique-width at most $k$ can be solved in ${\cal O}((kn)^2)$ time~\cite{kratsch_et_al:LIPIcs:2020:11899}. As a by-product of our Theorem~\ref{thm:dls-new}, we observe below that we can improve the dependency on $k$, but at the price of a logarithmic overhead in the running time.

\begin{corollary}\label{cor:apsp}
For every $n$-vertex graph $G=(V,E)$, if $cw(G) \leq k$ then we can solve All-Pairs Shortest-Paths for $G$ in ${\cal O}(kn^2\log{n})$ time.
\end{corollary}

\begin{proof}
We start applying Theorem~\ref{thm:dls-new} in order to compute a distance-labeling scheme with ${\cal O}(k\log{n})$ query time. Since every $n$-vertex graph has at most ${\cal O}(n^2)$ edges, it can be done in ${\cal O}(kn^2\log{n})$ time. Then, we consider all pairs $u,v \in V$ and we compute $d_G(u,v)$ in ${\cal O}(k\log{n})$ time. 
\end{proof}

\section{The diameter problem and beyond}\label{sec:diam}

Given a graph $G=(V,E)$ and a vertex $v$, the eccentricity of $v$, denoted $e_G(v)$, is equal to $\max_{u \in V} d_G(u,v)$. In particular, we could define the diameter of $G$ as $diam(G) = \max_{v \in V} e_G(v)$.
We refine our strategy for the above Theorem~\ref{thm:dls-new} in order to prove the main result of this paper:

\begin{theorem}\label{thm:ecc}
For every connected $n$-vertex $m$-edge graph $G=(V,E)$, if $cw(G) \leq k$ and a corresponding $k$-expression is given, then we can compute all the eccentricities in ${\cal O}(2^{{\cal O}(k)}(n+m)^{1+o(1)})$ time. In particular, we can compute $diam(G)$ in ${\cal O}(2^{{\cal O}(k)}(n+m)^{1+o(1)})$ time. 
\end{theorem}

The total distance of a vertex $v$ is equal to $TD_G(v) = \sum_{u \in V} d_G(u,v)$. The Wiener index of $G$ is equal to $W(G) = \sum_{v \in V} TD_G(v)$. The median set of $G$ contains all vertices $v$ s.t. $TD_G(v)$ is minimized. With a similar proof as for Theorem~\ref{thm:ecc}, we get:

\begin{theorem}\label{thm:td}
For every connected $n$-vertex $m$-edge graph $G=(V,E)$, if $cw(G) \leq k$ and a corresponding $k$-expression is given, then we can compute all the total distances in ${\cal O}(2^{{\cal O}(k)}(n+m)^{1+o(1)})$ time. In particular, we can compute $W(G)$ and the median set of $G$ in ${\cal O}(2^{{\cal O}(k)}(n+m)^{1+o(1)})$ time. 
\end{theorem}

Recall that Coudert et al. proved that assuming SETH, for any $\epsilon > 0$, there is no  ${\cal O}(2^{o(k)}(n+m)^{2-\epsilon})$-time algorithm for computing the diameter within cubic graphs of clique-width at most $k$~\cite{CDP19}. They also observed that since the pathwidth of a graph is an upper bound for its clique-width~\cite{FRRS09}, then it follows from~\cite{AVW16} that it is already ``SETH-hard'' to decide whether the diameter is either two or three. It is well-known that $diam(G) \leq 2$ if and only if for every $v \in V$ of degree $d_G(v)$, $TD_G(v) = 2(n-1) - d_G(v)$~\cite{BHM20}. In particular, $diam(G) \leq 2$ if and only if $W(G) = 2n(n-1) - 2m$. As a result, our Theorem~\ref{thm:td} for the Wiener index is also optimal under SETH. 

\paragraph{Additional notations.} From this point on we need to consider {\em weighted graphs}, due to some technicalities in our final proof of Theorems~\ref{thm:ecc} and~\ref{thm:td}. For a weighted graph $G=(V,E,w)$, we call a cut $(A,V \setminus A)$ {\em unweighted} if all edges between $A$ and $V \setminus A$ have a unit weight. The neighbourhood diversity of a cut is the same in $G$ as in the underlying unweighted graph obtained from $G$ by replacing all the weights by $1$. Similarly, a $k$-module of $G$ is a $k$-module in its underlying unweighted graph.

\subsection{Minimal partition of $k$-modules}\label{sec:k-mod}

First, it is not hard to show that every $k$-module has a partition in a least number of subsets. In what follows, we will often use a few simple properties of this minimal partitioning.

\begin{lemma}\label{lem:canonical-k-mod}
Every vertex-subset $A$ in a graph $G=(V,E,w)$ admits a unique partition $A_1,A_2,\ldots,A_k$ with the following two properties:
\begin{enumerate}
\item For every $1 \leq i \leq k$, for every $u_i,v_i \in A_i$, we have $N_G(u_i) \setminus A = N_G(v_i) \setminus A$. In particular, $A$ is a $k$-module of $G$.
\item For every $k' < k$, $A$ is {\em not} a $k'$-module of $G$.
\end{enumerate}
We call it the minimal partition of $A$, and it can be computed in linear time. 
\end{lemma}

\begin{proof}
Let $G' = G \setminus E(A)$ be the graph obtained from $G$ by removing all edges with their two ends in $A$. Two vertices are called false twins if they have exactly the same neighbours in $G'$. This is an equivalence relation over $V$, whose equivalence classes are sometimes called ``twin classes''. We claim that if $A$ is a $k'$-module, with a corresponding partition being $A_1,A_2,\ldots,A_{k'}$, then for every $1 \leq i \leq k'$, all the vertices of $A_i$ must belong to the same twin class. Indeed, for every $u_i,v_i \in A_i$ we get $N_{G'}(u_i) = N_G(u_i) \setminus A = N_G(v_i) \setminus A = N_{G'}(v_i)$. Then, the minimal partition of $A$ is composed of all the non-empty intersections of $A$ with the twin classes of $G'$. The twin classes of a graph can be computed in linear time by using classic partition refinement techniques~\cite{HMPV00}. 
\end{proof}

\subsection{Orthogonal range queries}\label{sec:rq}

We then need to recall some basics about the framework introduced in~\cite{CaK09} by Cabello and Knauer.
Let $P \subseteq \mathbb{R}^k$ be a static set of $k$-dimensional points. We assume each point $\overrightarrow{p} \in P$ to be assigned a value $g(\overrightarrow{p})$. A box is the Cartesian product of $k$ intervals. Note that we allow each interval to be unbounded and/or open or partially open. Roughly, given a box ${\cal R}$, a range query on $P$ asks for either reporting or counting all points in $P \cap {\cal R}$, or for some specific point(s) in this intersection maximizing a given objective function. Here, we consider the following types of range queries:
\begin{itemize}
\item({\it Maximum range query}) Given some box ${\cal R}$, find some $\overrightarrow{p} \in P \cap {\cal R}$ maximizing $g(\overrightarrow{p})$;
\item({\it Sum range query}) Given some box ${\cal R}$, compute $\sum_{\overrightarrow{p} \in P \cap {\cal R}}g(\overrightarrow{p})$.
\item({\it Count range query}) Given some box ${\cal R}$, compute $|P \cap {\cal R}|$.
\end{itemize}

\begin{lemma}[\cite{BHM20}]\label{lem:range-tree}
For every $k$-dimensional point set $P$ of size $n$, we can construct in ${\cal O}(2^{{\cal O}(k)} n^{1+o(1)})$ time a data structure, sometimes called a $k$-dimensional range tree, that allows to answer any maximum range query, sum range query or count range query in ${\cal O}(2^{{\cal O}(k)}n^{o(1)})$ time. 
\end{lemma}

In the following Lemma~\ref{lem:cut-rq} we give a new simple application of Lemma~\ref{lem:range-tree} to distance problems in graphs, namely: 

\begin{lemma}\label{lem:cut-rq}
Let $G=(V,E,w)$ be a connected $n$-vertex $m$-edge graph, let $(A,V \setminus A)$ be an unweighted cut of neighbourhood diversity at most $k$, and let $A' \subseteq A, \ B' \subseteq V \setminus A$. 
After a pre-processing in ${\cal O}(km + 2^{{\cal O}(k)} n^{1+o(1)})$ time, for every vertex $u \in A'$ we can compute the values $\max_{v \in B'} d_G(u,v)$ and $\sum_{v \in B'} d_G(u,v)$ in ${\cal O}(2^{{\cal O}(k)}n^{o(1)})$ time; in the same way, for every vertex $v \in B'$ we can compute the values $\max_{u \in A'} d_G(v,u)$ and $\sum_{u \in A'} d_G(v,u)$ in ${\cal O}(2^{{\cal O}(k)}n^{o(1)})$ time. 
\end{lemma}

\begin{proof}
Let $A_1,A_2,\ldots,A_k$ be the minimal partition of $A$. By Lemma~\ref{lem:canonical-k-mod}, we can compute it in ${\cal O}(m)$ time. For $1 \leq i \leq k$, let $B_i = N_G(A_i) \setminus A$. Note that since the subsets $A_i$ are pairwise disjoint, we can compute $B_1,B_2,\ldots,B_k$ in total ${\cal O}(m)$ time.
Observe that there is at most one index $i$ s.t. $B_i = \emptyset$ (otherwise, we can merge all groups $A_j$ s.t. $B_j = \emptyset$ into one, thus contradicting the minimality of the partition of $A$). W.l.o.g., if such index exists then it must be $i=k$. We want to exclude this index, if it exists, in order to avoid handling with arithmetic over infinite values. So, let $k' = k$ if $B_k \neq \emptyset$, otherwise let $k' = k-1$.
For every $1 \leq i \leq k'$, for every $u \in A'$, we compute $d_G(u,A_i)$. In the same way, for every $1 \leq i \leq k'$, for every $v \in B'$, we compute $d_G(B_i,v)$. It takes ${\cal O}(k'm) = {\cal O}(km)$ time in total if we use the single-source shortest-path algorithm of Thorup~\cite{thorup1999undirected}. Then, for every $v \in B'$ and for every $1 \leq i \leq k'$, we create a $k'$-dimensional point $\overrightarrow{p}(v,i)$: whose first coordinate is the index $i$, followed by the values $d_G(B_i,v) - d_G(B_j,v), \ 1 \leq j \leq k', \ j \neq i$. Set $g(\overrightarrow{p}(v,i)) = d_G(B_i,v)$. Finally, let $P$ contain all these $k'|B'|$ points.  We add all points in $P$ into some $k'$-dimensional range tree, that takes ${\cal O}(2^{{\cal O}(k)} n^{1+o(1)})$ time by Lemma~\ref{lem:range-tree}. 

Now, let $u \in A'$ be fixed, and assume that we want to compute the values $\max_{v \in B'} d_G(u,v)$ and $\sum_{v \in B'} d_G(u,v)$. By Lemma~\ref{lem:dist-formula}, for every $v \in B'$, we have $d_G(u,v) = \min\{ d_G(u,A_i) + 1 + d_G(B_i,v) \mid 1 \leq i \leq k \}$. Since $d_G(B_i,v) = +\infty$ if $B_i = \emptyset$, we also have $d_G(u,v) = \min\{ d_G(u,A_i) + 1 + d_G(B_i,v) \mid 1 \leq i \leq k' \}$. We (virtually) partition $B'$ into $C_1, C_2, \ldots C_{k'}$ so that, for every $1 \leq i \leq k'$, $v \in C_i$ if and only if the least index $j$ s.t. $d_G(u,v) = d_G(u,A_j) + 1 + d_G(B_j,v)$ is equal to $i$. Specifically, we design boxes ${\cal R}_1, {\cal R}_2, \ldots, {\cal R}_{k'}$ so that $p(v,j) \in {\cal R}_i \Longleftrightarrow j = i \ \text{and} \ v \in C_i$. Note that if we can do so, then:
\begin{align*}
\max_{v \in B'} d_G(u,v) &= \max_{1 \leq i \leq k'} \max_{v \in C_i} d_G(u,v) \\
&= \max_{1 \leq i \leq k'} \left( d_G(u,A_i) + 1 + \max\{ d_G(B_i,v) \mid v \in C_i  \} \right) \\
&= \max_{1 \leq i \leq k'} \left( d_G(u,A_i) + 1 + \max\{ g(\overrightarrow{p}(v,j)) \mid  \overrightarrow{p}(v,j) \in {\cal R}_i \} \right).
\end{align*}
In particular, we are left doing $k'$ maximum range queries. In the same way:
\begin{align*}
\sum_{v \in B'} d_G(u,v) &= \sum_{i=1}^{k'} \sum_{v \in C_i} d_G(u,v) \\
&= \sum_{i=1}^{k'} \sum_{v \in C_i} \left( d_G(u,A_i) + 1 + d_G(B_i,v) \right) \\
&= \sum_{i=1}^{k'} \left[  \left(d_G(u,A_i) + 1 \right) \cdot |C_i| + \sum_{v \in C_i} d_G(B_i,v)  \right] \\
&= \sum_{i=1}^{k'} \left[  \left(d_G(u,A_i) + 1 \right) \cdot |P \cap {\cal R}_i| + \sum \left\{ g(\overrightarrow{p}(v,j)) \mid \overrightarrow{p}(v,j) \in {\cal R}_i \right\}\right].
\end{align*}
In particular, we are left doing $k'$ sum range queries and $k'$ count range queries. Hence, being given ${\cal R}_1, {\cal R}_2, \ldots, {\cal R}_{k'}$, we are done in ${\cal O}(2^{{\cal O}(k)}n^{o(1)})$ time by Lemma~\ref{lem:range-tree}.

For every $1 \leq i \leq k'$, the box ${\cal R}_i$ is defined as follows. Let $\overrightarrow{p} = (p_1,p_2,\ldots,p_{k'})$ be a $k'$-dimensional point. We have $\overrightarrow{p} \in {\cal R}_i$ if and only if:
$$\begin{cases}
p_1 = i \\
\forall 1 \leq j \leq i-1, \ p_{j+1} < \left(d_G(u,A_j) - d_G(u,A_i)\right)  \\
\forall i+1 \leq j \leq k', \ p_j \leq  \left(d_G(u,A_j) - d_G(u,A_i)\right).
\end{cases}$$
Indeed, we have:
\begin{align*}
d_G(u,A_i) + 1 + d_G(B_i,v) \leq d_G(u,A_j) + 1 + d_G(B_j,v) 
&\Longleftrightarrow d_G(u,A_i) + d_G(B_i,v) \leq d_G(u,A_j) + d_G(B_j,v) \\
&\Longleftrightarrow d_G(B_i,v) - d_G(B_j,v) \leq \left(d_G(u,A_j) - d_G(u,A_i)\right).
\end{align*}
Furthermore, by construction, if $j < i$ then $d_G(B_i,v) - d_G(B_j,v)$ is exactly the $(j+1)^{th}$ coordinate of $\overrightarrow{p}(v,i)$ (in which case we want the inequality to be strict by the definition of $C_i$), and if $j > i$ then $d_G(B_i,v) - d_G(B_j,v)$ is exactly the $j^{th}$ coordinate of this point.

\medskip
For the vertices $v \in B'$, we proceed similarly as above, that is, we create a point-set $P'$ from the vertices in $A'$ and we put them in some separate $k'$-dimensional range tree.
\end{proof}

\subsection{Distance-preservers with weighted edges}\label{sec:gadget}

Our next objective consists in adding some weighted subsets to the two sides of a cut in order to preserve the distances from the original graph. Recall that for every two subsets $U$ and $W$, $d_G(U,W) = \min_{u \in U, w \in W} d_G(u,w)$. Our construction below is inspired by Cunningham's split decomposition~\cite{Cun82}.

\begin{definition}\label{def:gadget}
Given $G=(V,E,w)$ connected, let $(A, V \setminus A)$ be an unweighted cut of neighbourhood diversity at most $k$. Let $A_1,A_2,\ldots,A_k$ be the minimal partition of $A$. W.l.o.g., either all the $B_i$'s are nonempty, or $B_k$ is the unique empty set amongst the $B_i$'s. We set $k'=k$ if $B_k \neq \emptyset$, and $k' = k-1$ otherwise.
\begin{itemize}
\item The graph $H_A$ is obtained from $G[A]$ by the addition of $(k')^2$ fresh new vertices $b_{ij}, \ 1 \leq i,j \leq k'$. For every $1 \leq i \leq k'$, we add an edge of unit weight between every vertex $b_{ij}$ and every vertex of $A_i$. For every $1 \leq i < j \leq k'$, we add an edge $b_{ij}b_{ji}$ of weight $d_G(B_i,B_j)$. 
\item The graph $H_B$ is obtained from $G \setminus A$ by the addition of $(k')^2$ fresh new vertices $a_{ij}, \ 1 \leq i,j \leq k'$. For every $1 \leq i \leq k'$, we add an edge of unit weight between every vertex $a_{ij}$ and every vertex of $B_i$. For every $1 \leq i < j \leq k'$, we add an edge $a_{ij}a_{ji}$ of weight $d_G(A_i,A_j)$. 
\end{itemize}
\end{definition}

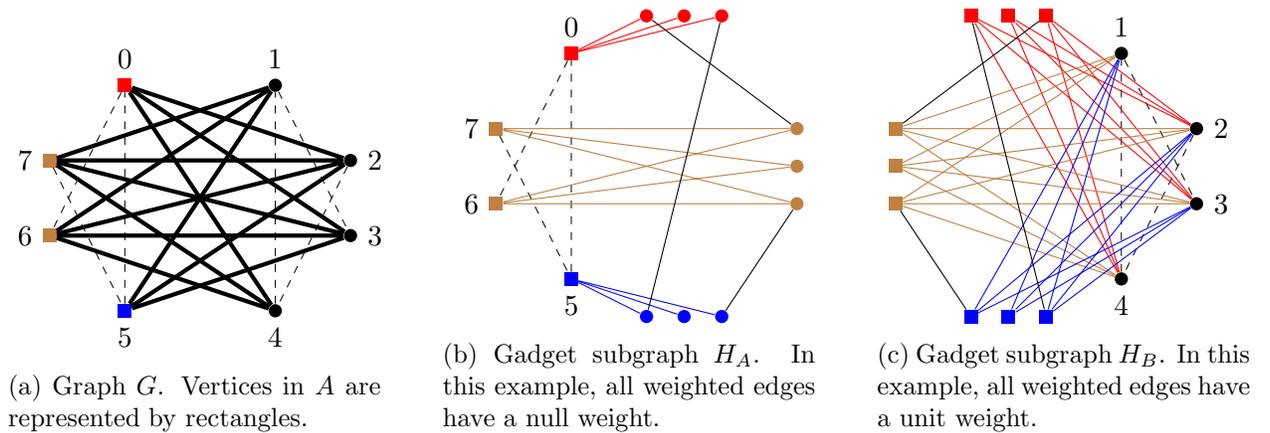
\begin{figure}[!h]
	\begin{center}
		\begin{subfigure}[b]{.3\textwidth}\centering
			\begin{tikzpicture}
		
				\node[rectangle,fill=red,inner sep=0pt,minimum size=5pt,label=above:{$0$}] (0) at (-1, 0) {};
				\node[circle,fill=black,inner sep=0pt,minimum size=5pt,label=above:{$1$}] (1) at (1, 0) {};	
				\node[circle,fill=black,inner sep=0pt,minimum size=5pt,label=right:{$2$}] (2) at (2, -1) {};
				\node[circle,fill=black,inner sep=0pt,minimum size=5pt,label=right:{$3$}] (3) at (2, -2) {};
				\node[circle,fill=black,inner sep=0pt,minimum size=5pt,label=below:{$4$}] (4) at (1, -3) {};
				\node[rectangle,fill=blue,inner sep=0pt,minimum size=5pt,label=below:{$5$}] (5) at (-1, -3) {};	
				\node[rectangle,fill=brown,inner sep=0pt,minimum size=5pt,label=left:{$6$}] (6) at (-2, -2) {};
				\node[rectangle,fill=brown,inner sep=0pt,minimum size=5pt,label=left:{$7$}] (7) at (-2, -1) {};			
		
				\draw[ultra thin,dashed] (7) -- (5) -- (0) -- (6);
				\draw[ultra thin,dashed] (2) -- (4) -- (1) -- (3);
				\draw[ultra thick] (2) -- (0) -- (3) -- (5) -- (2) -- (7) -- (3) -- (6) -- (2);
				\draw[ultra thick] (0) -- (4) -- (6) -- (1) -- (7) -- (4);
				\draw[ultra thick] (1) -- (5);
				
			\end{tikzpicture}
			\caption{Graph $G$. Vertices in $A$ are represented by rectangles.}
		\end{subfigure}\hfill
		\begin{subfigure}[b]{.3\textwidth}\centering
			\begin{tikzpicture}
			
				\node[rectangle,fill=red,inner sep=0pt,minimum size=5pt,label=above:{$0$}] (0) at (-1, 0) {};
				\node[rectangle,fill=blue,inner sep=0pt,minimum size=5pt,label=below:{$5$}] (5) at (-1, -3) {};	
				\node[rectangle,fill=brown,inner sep=0pt,minimum size=5pt,label=left:{$6$}] (6) at (-2, -2) {};
				\node[rectangle,fill=brown,inner sep=0pt,minimum size=5pt,label=left:{$7$}] (7) at (-2, -1) {};			
				\draw[ultra thin,dashed] (7) -- (5) -- (0) -- (6);

				\node[circle,fill=red,inner sep=0pt,minimum size=5pt] (a) at (0,.5) {};	
				\node[circle,fill=red,inner sep=0pt,minimum size=5pt] (b) at (.5,.5) {};	
				\node[circle,fill=red,inner sep=0pt,minimum size=5pt] (c) at (1,.5) {};	
				
				\draw[thin,red] (a) -- (0) -- (c);
				\draw[thin,red] (0) -- (b);			
				
				\node[circle,fill=brown,inner sep=0pt,minimum size=5pt] (d) at (2,-1) {};
				\node[circle,fill=brown,inner sep=0pt,minimum size=5pt] (e) at (2,-1.5) {};	
				\node[circle,fill=brown,inner sep=0pt,minimum size=5pt] (f) at (2,-2) {};
				
				\draw[thin,brown] (6) -- (d) -- (7) -- (f) -- (6) -- (e) -- (7);				
				
				\node[circle,fill=blue,inner sep=0pt,minimum size=5pt] (g) at (0,-3.5) {};
				\node[circle,fill=blue,inner sep=0pt,minimum size=5pt] (h) at (.5,-3.5) {};		
				\node[circle,fill=blue,inner sep=0pt,minimum size=5pt] (i) at (1,-3.5) {};
				
				\draw[thin,blue] (g) -- (5) -- (i);
				\draw[thin,blue] (h) -- (5);

				\draw (a) -- (d);
				\draw (c) -- (g);
				\draw (f) -- (i);

			\end{tikzpicture}
			\caption{Gadget subgraph $H_A$. In this example, all weighted edges have a null weight.}
		\end{subfigure}\hfill
		\begin{subfigure}[b]{.3\textwidth}\centering
			\begin{tikzpicture}
			
				\node[circle,fill=black,inner sep=0pt,minimum size=5pt,label=above:{$1$}] (1) at (1, 0) {};	
				\node[circle,fill=black,inner sep=0pt,minimum size=5pt,label=right:{$2$}] (2) at (2, -1) {};
				\node[circle,fill=black,inner sep=0pt,minimum size=5pt,label=right:{$3$}] (3) at (2, -2) {};
				\node[circle,fill=black,inner sep=0pt,minimum size=5pt,label=below:{$4$}] (4) at (1, -3) {};		
		
				\draw[ultra thin,dashed] (2) -- (4) -- (1) -- (3);

				\node[rectangle,fill=red,inner sep=0pt,minimum size=5pt] (a) at (0,.5) {};
				\node[rectangle,fill=red,inner sep=0pt,minimum size=5pt] (b) at (-.5,.5) {};		
				\node[rectangle,fill=red,inner sep=0pt,minimum size=5pt] (c) at (-1,.5) {};	
			
				\draw[thin,red] (c) -- (2) -- (a) -- (3) -- (c) -- (4) -- (a);
				\draw[thin,red] (2) -- (b) -- (3);
				\draw[thin,red] (b) -- (4);			
				
				\node[rectangle,fill=brown,inner sep=0pt,minimum size=5pt] (d) at (-2,-1) {};
				\node[rectangle,fill=brown,inner sep=0pt,minimum size=5pt] (e) at (-2,-1.5) {};	
				\node[rectangle,fill=brown,inner sep=0pt,minimum size=5pt] (f) at (-2,-2) {};
								
				\draw[thin,brown] (d) -- (1) -- (f) -- (2) -- (d) -- (3) -- (f) -- (4) -- (d);	
				\draw[thin,brown] (1) -- (e) -- (2);
				\draw[thin,brown] (3) -- (e) -- (4);			
				
				\node[rectangle,fill=blue,inner sep=0pt,minimum size=5pt] (g) at (0,-3.5) {};
				\node[rectangle,fill=blue,inner sep=0pt,minimum size=5pt] (h) at (-.5,-3.5) {};	
				\node[rectangle,fill=blue,inner sep=0pt,minimum size=5pt] (i) at (-1,-3.5) {};
				
				\draw[thin,blue] (i) -- (2) -- (g) -- (3) -- (i) -- (1) -- (g);
				\draw[thin,blue] (2) -- (h) -- (3);
				\draw[thin,blue] (h) -- (1);		
				
				\draw (a) -- (d);
				\draw (c) -- (g);
				\draw (f) -- (i);
			\end{tikzpicture}
			\caption{Gadget subgraph $H_B$. In this example, all weighted edges have a unit weight.}
		\end{subfigure}
		\caption{An illustration of the procedure of Definition~\ref{def:gadget}.}
		\label{fig:gadget}
	\end{center}
\end{figure}

We refer to Fig.~\ref{fig:gadget} for a illustration. Observe that in both gadget subgraphs $H_A$ and $H_B$, the newly added vertices induce a matching of cardinality $k'(k'-1)/2$, where $k' \in \{k-1,k\}$.
Below, we observe that it is rather straightforward to compute these two above subgraphs $H_A$ and $H_B$ in parameterized linear time:

\begin{lemma}\label{lem:construct-gadget}
Given $G=(V,E,w)$ connected, let $(A, V \setminus A)$ be an unweighted cut of neighbourhood diversity at most $k$. 
The gadget subgraphs $H_A$ and $H_B$ (see Definition~\ref{def:gadget}) can be constructed in ${\cal O}(k^2n + km)$ time.
\end{lemma}

\begin{proof}
Since, in both $G[A]$ and $G \setminus A$, we only add ${\cal O}(k^2)$ new vertices, there are ${\cal O}(k^2n)$ new edges to create. Each edge can be created in ${\cal O}(1)$ time if for all $1 \leq i \leq k$, the subsets $A_i,B_i$, as they were defined in Definition~\ref{def:gadget}, are given. As we already observed in the proof of Lemma~\ref{lem:cut-rq}, these $2k$ subsets can be created in ${\cal O}(m)$ time by using partition refinement techniques. Then, it only remains to compute the edge-weights. For every fixed $i$, we can compute $d_G(A_i,A_j)$ for every $1 \leq j \leq k, \ j \neq i$, as follows. We compute $d_G(v,A_i)$ for every $v \in V$. It takes ${\cal O}(m)$ time if we use the single-source shortest-path algorithm of Thorup~\cite{thorup1999undirected}. Then, in additional ${\cal O}(n)$ time we scan each subset $A_j$, and we keep a $v \in A_j$ minimizing $d_G(v,A_i)$. We do the same in order to compute the distances $d_G(B_i,B_j)$. However, since now the subsets $B_j$ may not be disjoint, the running time goes up to ${\cal O}(m+kn)$ for every fixed $i$. 
\end{proof}

The following two properties are crucial in our proofs of Theorems~\ref{thm:ecc} and~\ref{thm:td}. First, we prove that our gadget subgraphs effectively ``repair'' the distances in the two subgraphs resulting from a cut, making them coincide with the distances in $G$.

\begin{lemma}\label{lem:dist-preserver}
Given $G=(V,E,w)$ connected, let $(A, V \setminus A)$ be an unweighted cut of neighbourhood diversity at most $k$. Let $H_A,H_B$ be as in Definition~\ref{def:gadget}. Then, for every $u,v \in A$ we have $d_G(u,v) = d_{H_A}(u,v)$. Similarly, for every $u,v \notin A$ we have $d_G(u,v) = d_{H_B}(u,v)$.
\end{lemma}

\begin{proof}
We only detail the proof for $u,v \in A$. First, we prove that $d_{H_A}(u,v) \leq d_G(u,v)$. Indeed, if there exists a $uv$-path of weight $d_G(u,v)$ which is fully into $A$, then this path also exists in $H_A$. Otherwise, every shortest $uv$-path in $G$ must intersect $V \setminus A$. Let us fix a shortest $uv$-path $P$ in $G$. We scan $P$ from $u$ until we find the first edge $xy$ s.t. $x \in A, y \notin A$. Similarly, we scan $P$ from $v$ until we find the first edge $st$ s.t. $s \in A, \ t \notin A$. There exist $i,j$ s.t. $x \in A_i, s \in A_j$, and so, $y \in B_i, t \in B_j$. We have $d_G(y,t) \geq d_G(B_i,B_j)$, and this is in fact an equality because $P$ is a shortest $uv$-path and there are complete joins between $A_i$ and $B_i$, respectively between $A_j$ and $B_j$. Then, we may replace all the $yt$-subpath in $P$ by either the edge $b_{ij}b_{ji}$ (if $i \neq j$), or simply $b_{ii}$ (if $i = j$). Doing so, we obtain a $uv$-path of $H_A$ of weight equal to $d_G(u,v)$. Conversely, we prove that $d_{H_A}(u,v) \geq d_G(u,v)$. Indeed, consider any $uv$-path $P'$ of $H_A$. If $P' \subseteq A$ then it is also a $uv$-path in $G$. Otherwise, we scan $P'$ from $u$ until we find a vertex $b_{ij}$. There are two cases:
\begin{itemize}
\item \underline{Case 1}: the next vertex onto $P'$ is some vertex $y \in A$. Let also $x \in A$ be the predecessor of $b_{ij}$ onto $P'$ (neighbour of $b_{ij}$ onto the subpath of $P'$ between $u$ and $b_{ij}$). Observe that $x,y \in A_i$. Furthermore, $B_i \neq \emptyset$ (otherwise, according to the process of Definition~\ref{def:gadget}, the vertex $b_{ij}$ would not exist). We replace $b_{ij}$ by any vertex of $B_i$, then we continue scanning the sub-path of $P'$ between $y$ and $v$.
\item \underline{Case 2}: the next vertex onto $P'$ is $b_{ji}$.  Let $x,y \in V(P')$ be respectively the predecessor of $b_{ij}$ and successor of $b_{ji}$ (starting from $u$). Note that $x,y \in A$. Furthermore, we stress that $B_i,B_j$ are nonempty. We replace the edge $b_{ij}b_{ji}$ by any shortest $B_iB_j$-path in $G$, then we continue scanning the sub-path of $P'$ between $y$ and $v$.
\end{itemize}
In both cases, we transform $P'$ into a $uv$-path of $G$ without changing the weight. Finally, the proof for $u,v \notin A$ is similar as what we did above, and in fact it is a bit simpler because we can {\em never} have $A_i = \emptyset$ for any $i$.
\end{proof}

Our approach only works for unweighted cuts. In particular, if we want to apply the procedure of Definition~\ref{def:gadget} recursively, for some cuts in the gadget subgraphs $H_A$ and $H_B$, then we must have all the new vertices in these subgraphs ({\it i.e.}, those incident to weighted edges) on a same side of the cut. The next lemma shows that restricting ourselves to such cuts does not cause an explosion of their neighbourhood diversity.

\begin{lemma}\label{lem:cut-preserver}
Given $G=(V,E,w)$ connected, let $(A, V \setminus A)$ be an unweighted cut of neighbourhood diversity at most $k$. Let $H_A,H_B$ be as in Definition~\ref{def:gadget}.
\begin{enumerate}
\item For every $A' \subseteq A$, if $A'$ is a $k$-module of $G$ then it is a $k$-module of $H_A$.
\item For every $B' \subseteq V \setminus A$, if $B'$ is a $k$-module of $G$ then it is a $k$-module of $H_B$; if $A \cup B'$ is a $k$-module of $G$ then $B' \cup \{ a_{ij} \mid 1 \leq i,j \leq k' \}$ is a $k$-module of $H_B$. 
\end{enumerate}

\begin{proof}
Let $A_1,A_2,\ldots,A_k$ and $B_1, B_2, \ldots, B_k$ be as in Definition~\ref{def:gadget}.
By minimality of the partition of $A$, there are no two indices $i$ and $j$ s.t. $A_i \cup A_j$ is a module of $G \setminus (A \setminus (A_i \cup A_j))$ (otherwise, we could have merged these two groups into one).
We prove the properties of the lemma separately.
\begin{itemize}
\item Let us first assume that $A' \subseteq A$ is a $k$-module of $G$. Let $u,v \in A'$ be s.t. $N_G(u) \setminus A' = N_G(v) \setminus A'$. Since we have $N_G(u) \setminus A = N_G(v) \setminus A$, we must have that there exists a $j$ s.t. $u,v \in A_j$. In particular, $N_G(u) \setminus A = N_G(v) \setminus A = B_j$.
Then, $$N_{H_A}(u) \setminus A = N_{H_A}(v) \setminus A = \begin{cases} \{ b_{ji} \mid 1 \leq i \leq k', \ i \neq j \} \ \text{if} \ B_j \neq \emptyset \\
\emptyset \ \text{otherwise}.
\end{cases}$$
As a result, $N_{H_A}(u) \setminus A' = N_{H_A}(v) \setminus A'$, that proves that $A'$ is also a $k$-module of $H_A$. 
\item In the same way, let us now assume that $B' \subseteq V \setminus A$ is a $k$-module of $G$. Let $u,v \in B'$ be s.t. $N_G(u) \setminus B' = N_G(v) \setminus B'$. In particular, $N_G(u) \cap A = N_G(v) \cap A$, and so, for every $1 \leq i \leq k'$, $u \in B_i \Longleftrightarrow v \in B_i$. It implies $N_{H_B}(u) \cap \{ a_{ij} \mid 1 \leq i,j \leq k' \} = N_{H_B}(v) \cap \{ a_{ij} \mid 1 \leq i,j \leq k' \}$. Since in addition, $N_G(u) \setminus (A \cup B') = N_G(v) \setminus (A \cup B')$, we get $N_{H_B}(u) \setminus B' = N_{H_B}(v) \setminus B'$. As a result, $B'$ is also a $k$-module of $H_B$. 
\item Finally, let $B' \subseteq V \setminus A$ be s.t. $B' \cup A$ is a $k$-module of $G$. In particular, let $C_1,C_2,\ldots,C_k$ be a corresponding $k$-partition of $B' \cup A$. Write $B'' = B' \cup \{ a_{ij} \mid 1 \leq i,j \leq k' \}$. We observe that for every $1 \leq i \leq k'$, for all $x_i,y_i \in A_i$ we have $N_G(x_i) \setminus (B' \cup A) = N_G(y_i) \setminus (B' \cup A)$. Hence, we may assume the existence of some index $p$ s.t. $A_i \subseteq C_p$. Furthermore, for any $j$, $N_{H_B}(a_{ij}) \setminus B'' = B_i \setminus B' = N_G(x_i) \setminus (B' \cup A)$. Then, for every $1 \leq p \leq k$, we define $C_p'$ as containing $C_p \setminus A$ and, for every $i$ s.t. $A_i \subseteq C_p$, all the vertices $a_{ij}$. The partition $C_1',C_2',\ldots,C_k'$ certifies that $B''$ is indeed a $k$-module of $H_B$.
\end{itemize}
We stress that in contrast to the above, if $A' \cup (V \setminus A)$ is a $k$-module of $G$, then in general $A' \cup \{ b_{ij} \mid 1 \leq i,j \leq k' \}$ is {\em not} a $k$-module of $H_A$. Indeed, this is because the vertices in some subset $B_i$ may have different neighbourhoods in $A \setminus A'$.
\end{proof}

\end{lemma}

Although it is tempting to plug the procedure of Definition~\ref{def:gadget} in our construction of a distance-labeling scheme for bounded clique-width graphs (Sec.~\ref{sec:dls}), we observe that it would lead to a quadratic dependency on the clique-width in the running-time of Theorem~\ref{thm:dls-new}.

\subsection{Proofs of the main results}\label{sec:proof}

\begin{proof}[Proof of Theorems~\ref{thm:ecc} and~\ref{thm:td}]
We revisit the scheme of Theorem~\ref{thm:dls-new}. That is, we fix some width-$k$ partition tree $(T,f)$, that takes ${\cal O}(k(n+m))$ time by using Lemma~\ref{lem:partition-tree}. Furthermore, we pre-process the tree $T$ in order to compute in ${\cal O}(1)$ time, for any two nodes $a,a' \in V(T)$, their least common ancestor; it can be done in ${\cal O}(n)$ time~\cite{harel1984fast}. Finally, let $w : V(T) \to \{0,1\}$ be s.t. $w(a) = 1$ if and only if $a$ is a leaf. In what follows, we mimic the recursive construction of a $w$-centroid decomposition $T^{(w)}$ of $T$, as it was defined in the proof of Theorem~\ref{thm:dls-new}. 

\smallskip
{\bf The algorithm.}
We consider a more general problem for which we are given some tuple $\langle r, H, U, T^U, f^U, {\cal L} \rangle$. Let us detail each of the components of this input:
\begin{enumerate}
\item Here, $H$ is an edge-weighted graph with non-negative integer weights (initially, $H = G$).
\item The value $r$ represents the recursion level of the algorithm (initially, $r=0$).
\item The vertex-subset $U$ is such that $V \cap V(H) = U$ (initially, $U = V$). We further impose to have $H[U] = G[U]$, and that for every $u,v \in U$ we have $d_G(u,v) = d_H(u,v)$. In particular, all the edges of $H[U]$ have unit weight.
\item The rooted tree $(T^U,f^U)$ is a width-$k$-partition tree of $G[U]$ (initially, $T^U = T$ and $f^U = f$). We further assume that $T^U$ was constructed from a rooted subtree of $T$ by repeatedly removing internal nodes with only one child. In particular, all the ancestor-descendant relations in $T^U$ are also ancestor-descendant relations in $T$. Furthermore, for every node $b \in V(T^U)$ we impose $f^U(b) = \{ X \cap U \mid X \in f(b) \}$. Note that in lieu of $(T^U,f^U)$, we are given the representation graph of this partition tree (as defined in Sec.~\ref{sec:cut-tree}).
\item Finally, $H \setminus U$ is a disjoint union of $r' \leq r$ subgraphs of order ${\cal O}(k^2)$, that we shall name ``clusters'' in what follows. To each cluster $W_i$, we associate some node $c_i$ of the original tree $T$. Roughly, $c_i$ corresponds to some balanced cut, computed at an earlier recursive stage, and the cluster $W_i$ resulted from the procedure of Definition~\ref{def:gadget} applied to this cut. So, in particular, we impose that any edge between two vertices that are on different clusters (resp., between a vertex in a cluster and a vertex of $U$) must be unweighted. All the pairs $(W_i,c_i)$ are stored in the list ${\cal L}$ (initially, ${\cal L}$ is the empty list).
\end{enumerate}
The output of the algorithm is, for every $u \in U$, the values $\max_{v \in U} d_H(u,v)$ and $\sum_{v \in U} d_H(u,v)$. For that, let $n_r := |V(H)|$ and $m_r := |E(H)|$. We may assume that $|U| \geq \alpha k^2 \log{n}$, for some sufficiently large constant $\alpha$. Indeed, if it not the case then we may compute by brute-force all the desired values. Our algorithm has at most ${\cal O}(\log{n})$ recursive stages, and therefore, in this case we have $n_r = |U| + {\cal O}(k^2\log{n}) = {\cal O}(k^2\log{n})$. In particular, we can perform the brute-force computation in ${\cal O}(k^6\log^3{n})$ time (base case of the recursion). Thus from now on, let us assume $|U| = \Omega(k^2 \log{n})$. We compute a $w$-centroid $c$ in $T^U$. This can be done in ${\cal O}(|V(T^U)|) = {\cal O}(n_r)$ time. Since $w(T^U) = |U| > 3$, this node $c$ cannot be a leaf. Let $a_1,a_2,\ldots,a_d$ be the children of $c$. As before, we denote by $C$ (resp. $A_i$) the subset of vertices of which $f^U(c)$ (resp., $f^U(a_i)$) is a partition, and by $T^U_c$ (resp., $T^U_{a_i}$) the subtree rooted at $c$ (resp., at $a_i$). Here, we stress that $C \subseteq U$ (resp., $A_i \subseteq U$). By using Lemma~\ref{lem:bipartition-centroid}, we may partition $T^U \setminus \{c\}$ in two non-empty forests of respective weights $\leq 2|U|/3$. Furthermore, we may assume one of our two forests to contain exactly $T^U_{a_1},T^U_{a_2},\ldots,T^U_{a_p}$ for some $p \leq d$. Then, let $A = \bigcup_{j=1}^p A_j$ (computable in ${\cal O}(|U|) = {\cal O}(n_r)$ time by traversal of $T^U$). We compute the following cut of $H$:
\begin{itemize}
\item The subsets $A$ and $U \setminus A$ are on separate sides of the cut. 
\item For every $(W_j,c_j) \in {\cal L}$, there are two cases. If there exists some index $i$ s.t. the least common ancestor of $c_j$ and $a_i$ in $T$ is a {\em strict} descendant of $c$ (a child of $c$ in $T$, or a descendant of one of these children), then we put $W_j$ on the same side of the cut as $A$. Otherwise, we put $W_j$ on the same side of the cut as $U \setminus A$.
\end{itemize}
Note that, for each $(W_j,c_j) \in {\cal L}$, we can decide in which case we are as follows. For every $1 \leq i \leq p$, we compute the least common ancestor $s_i$ of $c_j$ and $a_i$ in $T$. Then, for every $1 \leq i \leq p$, we compute the least common ancestor of $s_i$ and $c$ in $T$. Given the pre-computed least-common ancestor data structure for $T$, this can be done in total ${\cal O}(p)$ time, and so in ${\cal O}(|U|) = {\cal O}(n_r)$ time. Overall, since we have $|{\cal L}| = r' = {\cal O}(\log{n})$, we can compute this above cut in ${\cal O}(n_r\log{n})$ time. Let $(A',V(H) \setminus A')$ be this cut, where $A \subseteq A'$. By construction, it is unweighted. We prove below (see the Correctness part of the proof) that $A'$ is a $k$-module of $H$. Then, we apply Lemma~\ref{lem:cut-rq} in order to compute, for every $u \in A$, the values $\max_{v \in U \setminus A} d_H(u,v)$ and $\sum_{v \in U \setminus A} d_H(u,v)$ (resp., for every $v \in U \setminus A$, the values $\max_{u \in A} d_H(v,u)$ and $\sum_{u \in A} d_H(v,u)$). It takes ${\cal O}(2^{{\cal O}(k)}(n_r+m_r)^{1+o(1)})$ time. 

\smallskip
We are left computing for every $u \in A$, the values $\max_{u' \in A} d_H(u,u')$ and $\sum_{u' \in A} d_H(u,u')$ (resp., for every $v \in U \setminus A$, the values $\max_{v' \in U \setminus A} d_H(v,v')$ and $\sum_{v' \in U \setminus A} d_H(v,v')$). For that, we construct the gadget subgraphs $H_A$ and $H_B$, as in Definition~\ref{def:gadget}. By Lemma~\ref{lem:construct-gadget}, it can be done in ${\cal O}(k^2n_r + km_r)$ time. Let $(T^A,f^A)$ and $(T^B,f^B)$ be width-$k$ partition trees of $G[A]$ and $G[U] \setminus A$. Recall (see the proof of Theorem~\ref{thm:dls-new}) that the trees $T^A$ and $T^B$ can be computed in ${\cal O}(|U|)$ time from $T^U$ as follows: we start with $T^U_c \setminus \left(\bigcup_{i > p} T^U_{a_i} \right)$ and $T^U \setminus \left(\bigcup_{i=1}^p T^U_{a_i} \right)$, then we remove useless leaves and/or we repeatedly contract internal nodes with only one child. The corresponding partition function $f^A$, resp. $f^B$, is obtained from $f^U$ by removal in the partition at each node of all the vertices out of $A$, resp. by removal of all the vertices in $A$. Hence, being given the representation graph of $(T^U,f^U)$, the representation graphs of $(T^A,f^A)$ and $(T^B,f^B)$ can be computed in ${\cal O}(k|U|)$ time. Let ${\cal L}_A$ contain every $(W_j,c_j) \in {\cal L}$ s.t. $W_j \subseteq A'$; we also add in ${\cal L}_A$ a new cluster $(V(H_A) \setminus A',c)$. In the same way, let ${\cal L}_B$ contain every $(W_j,c_j) \in {\cal L}$ s.t. $W_j \subseteq V(H) \setminus A'$; we also add in ${\cal L}_B$ a new cluster $(V(H_B) \setminus V(H),c)$. We end up calling our algorithm recursively for the inputs $\langle r+1, H_A, A, T^A, f^A, {\cal L}_A \rangle$ and $\langle r+1, H_B, U \setminus A, T^B, f^B, {\cal L}_B \rangle$. 
 
\paragraph{Correctness.} There are two properties to check in order to prove the validity of our approach. The first such property is that, being given the two gadget subgraphs $H_A$ and $H_B$ resulting from $H$, the distances in $H$ (and so, in $G$) are preserved. This follows from Lemma~\ref{lem:dist-preserver}. The second property to be checked is that we always compute a cut $(A',V(H) \setminus A')$ of neighbourhood diversity at most $k$. We prove it by induction on $r$. Specifically, we prove the following slightly stronger property.

\begin{property}\label{pty-cut}
{\it For every $\langle r, H, U, T^U, f^U, {\cal L} \rangle$, let $s_1,s_2,\ldots,s_q$ be children of some node $s$ in $T^U$. Let $S_i$ be the subset of $U$ of which $f(s_i)$ is a partition, and set $S = \bigcup_{i=1}^q S_i$. Finally, let $S'$ be the union of $S$ with all subsets $W_j$, for $(W_j,c_j) \in {\cal L}$, s.t. the least common ancestor in $T$ of $c_j$ and some node $s_i$ is a strict descendant of $s$. Then, $S'$ is a $k$-module of $H$.}  
\end{property}

If $r=0$ then, since ${\cal L} = \emptyset$, this directly follows from Lemma~\ref{lem:k-mod}. Let us assume the property to be true for $\langle r, H, U, T^U, f^U, {\cal L} \rangle$. In what follows, we analyse the cuts in $H_A$ and $H_B$, respectively.

\medskip
\noindent
({\it Gadget subgraph $H_A$}). Let $s_1,s_2,\ldots,s_q$ be children nodes of some $s$ in $T^A$. Observe that $T^A$ is a subtree of $T^U$ (equal to either $T^U_c \setminus \left( \bigcup_{i > p} T^U_{a_i} \right)$, or $T^U_{a_1}$ if $p=1$). In particular, $s_1,s_2,\ldots,s_q$ are also children nodes of $s$ in $T^U$. Let $W_{r'+1} := V(H_A) \setminus A'$ be the only cluster of ${\cal L}_A$ that is not contained in ${\cal L}$ (constructed using the procedure of Definition~\ref{def:gadget} in order to create $H_A$). We define $S_0'$ as the union of $S$ with all the clusters $W_j$, for $(W_j,c_j) \in {\cal L}$, s.t. the least common ancestor in $T$ of $c_j$ and some node $s_i$ is a strict descendant of $s$. By the induction hypothesis, $S_0'$ is a $k$-module of $H$. 

\begin{claim}\label{claim:a}
For every $(W_j,c_j) \in {\cal L}$, we have $W_j \subseteq S_0' \Longrightarrow (W_j,c_j) \in {\cal L}_A$.
\end{claim}
\begin{proofclaim}
Recall that if $W_j \subseteq S_0'$, then $c_j$ is a strict descendant of $s$. In particular, by the very construction of $T^A$, $c_j$ is a descendant of some $a_i$, for $1 \leq i \leq p$, unless maybe if $s = c$. Moreover if $s=c$, then the nodes $s_1,s_2,\ldots,s_q$ must be a subset of the nodes $a_1,a_2,\ldots,a_p$. Therefore, in both cases, there exists an $i$ s.t. $c_j$ and $a_i$ have a least common ancestor in $T$ which is a strict descendant of $c$.
\end{proofclaim}

Then, by Claim~\ref{claim:a}, $S_0' \subseteq A'$. By Lemma~\ref{lem:cut-preserver}, $S_0'$ is a $k$-module of $H_A$. Finally, since $(W_{r'+1},c) \in {\cal L}_A$ and {\em all} nodes of $T^A$ are descendants of $c$, we get $W_{r'+1} \not\subseteq S'$, and so, $S' = S_0'$.
   
\medskip
\noindent
({\it Gadget subgraph $H_B$}). Let $s_1,s_2,\ldots,s_q$ be children nodes of some $s$ in $T^B$. By construction, in $T^U$, the node $s$ is a common ancestor of all the nodes $s_1,s_2,\ldots,s_q$ (it may not be their father node since we possibly contracted internal nodes in order to create $T^B$). Let $W_{r'+1} := V(H_B) \setminus V(H)$ be the only cluster of ${\cal L}_B$ that is not contained in ${\cal L}$ (constructed using the procedure of Definition~\ref{def:gadget} in order to create $H_B$). In our analysis below, we will often use the following observation: when creating $T^B$ from $T^U \setminus \left(  \bigcup_{i=1}^p T^U_{a_i}\right)$ only two nodes may be removed, namely, $c$ (if it has at most one child left) or its father node in $T^U$ (if $c$ becomes a leaf and it has exactly one sibling in $T^U$). There are now two cases to be considered.  

\begin{figure}[!h]
	\begin{center}
		\begin{tikzpicture}
			
			\node[circle,fill=black,inner sep=0pt,minimum size=5pt,label=above:{$c$}] (c) at (0, 0) {};	
			\node[circle,fill=black,inner sep=0pt,minimum size=5pt] (ba) at (0, -1) {};	
			\node[circle,fill=black,inner sep=0pt,minimum size=5pt,label=right:{$a_{i'}$}] (a) at (1, -2) {};
			\node[circle,fill=black,inner sep=0pt,minimum size=5pt,label=left:{$s$}] (s) at (0, -2.5) {};
			\node[circle,fill=black,inner sep=0pt,minimum size=5pt] (bs) at (0, -3) {};	
			\node[circle,fill=black,inner sep=0pt,minimum size=5pt,label=left:{$s_i$}] (si) at (-1, -4) {};
			\node[circle,fill=black,inner sep=0pt,minimum size=5pt,label=right:{$c_j$}] (cj) at (1, -4) {};
		
			\draw[snake it] (c) -- (ba) -- (a);		
			\draw[snake it] (ba) -- (s) -- (bs);
			\draw[snake it] (si) -- (bs) -- (cj);			
			
		\end{tikzpicture}
	\end{center}
	\caption{To the proof of Theorems~\ref{thm:ecc} and~\ref{thm:td}.}
	\label{fig:proof}
\end{figure}
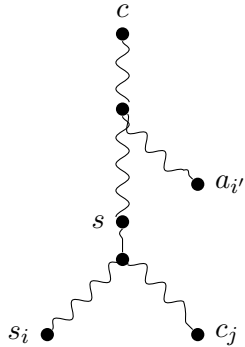

\begin{itemize}
\item We first assume that, for every $1 \leq i \leq q$, the least common ancestor of $c$ and $s_i$ is an ancestor of $s$ (possibly, $s$ itself). In particular, $f^B(s_i) = f^U(s_i)$. 

\begin{claim}\label{claim:b-1-1}
$s_1,s_2,\ldots,s_q$ are also children nodes of $s$ in $T^U$.
\end{claim}
\begin{proofclaim}
Suppose for the sake of contradiction that $s$ is not the father of $s_i$, for some $1 \leq i \leq q$. In particular, the original father node of $s_i$, let us call it $t_i$, got removed when we created $T^B$. But then, $t_i$ should be either $c$, or the father node of $c$ in $T^U$. As a result, $s_i$ and $c$ would have a least common ancestor in $T$ which is a strict descendant of $s$, a contradiction. 
\end{proofclaim}

The remainder of the proof is now essentially the same as what we did above for the gadget subgraph $H_A$. Specifically, let $S_0'$ be the union of $S$ with all the clusters $W_j$, for $(W_j,c_j) \in {\cal L}$, s.t. the least common ancestor in $T$ of $c_j$ and some node $s_i$ is a strict descendant of $s$. By the induction hypothesis, $S_0'$ is a $k$-module of $H$. Furthermore, the following result (similar to Claim~\ref{claim:a}) is true:

\begin{claim}\label{claim:b-1-2}
If $(W_j,c_j) \in {\cal L}$ is s.t. $W_j \subseteq S_0'$, then $(W_j,c_j) \in {\cal L}_B$.
\end{claim}
\begin{proofclaim}
Suppose for the sake of contradiction $(W_j,c_j) \in {\cal L}_A$. We refer to Fig.~\ref{fig:proof} for an illustration. In particular, for some $1 \leq i' \leq p$, $c_j$ and $a_{i'}$ have a common ancestor in $T$ which is a strict descendant of $c$. There also exists an $1 \leq i \leq q$ s.t. $c_j$ and $s_i$ have a common ancestor which is a strict descendant of $s$. Since both $s$ and $c$ are ancestors of $c_j$, one of these two nodes is an ancestor of the other. But $s$ cannot be a strict ancestor of $c$ (otherwise, the least common ancestor of $s_i$ and $c$ would be a strict descendant of $s$). Therefore, $c$ is an ancestor of $s$. Then, we consider two sub-cases. 
\begin{itemize}
\item First, let us assume $s = c$. Observe that $s_i \neq a_{i'}$ (otherwise, $s_i \notin V(T^B)$). Then, by Claim~\ref{claim:b-1-1}, $s_i$ and $a_{i'}$ are sibling nodes in $T^U$. Recall that the least common ancestor of $c_j$ and $s_i$ in $T$, resp. of $c_j$ and $a_{i'}$ in $T$, must be a strict descendant of $s=c$. As a result, the least common ancestor of $s_i$ and $a_{i'}$ in $T$, call it $t_i$, must be also a strict descendant of $c$ in $T$. This implies that $t_i$ got removed at some earlier recursive stage. But this is impossible, because at the stage when $t_i$ got removed, this node still had at least two children (being ancestors of $s_i$ and $a_{i'}$, respectively). A contradiction.
\item From now on, we assume $s \neq c$. We further observe that $a_{i'}$ cannot be a descendant of $s$ ({\it i.e.}, because $s \in V(T^U)$ and $a_{i'}$ is a child of $c$ in $T^U$). Therefore, the least common ancestor of $a_{i'}$ and $c_j$ should be on the $sc$-path in $T$. In fact, this least common ancestor must be $a_{i'}$ itself (otherwise, the least common ancestor of $s$ and $a_{i'}$ would be a strict descendant of $c$, that still exists in $T^U$ because it has at least two children, thus contradicting again that $a_{i'}$ is a child of $c$ in $T^U$). In particular, since $a_{i'}$ is onto the $sc$-path in $T$, $s \in T^U_{a_{i'}}$. But then, it contradicts our assumption that $s \in T^B$.   
\end{itemize}
Summarizing, in both sub-cases we derive a contradiction.
\end{proofclaim} 

By the above Claim~\ref{claim:b-1-2}, $S_0' \subseteq V(H) \setminus A'$. Hence, by Lemma~\ref{lem:cut-preserver}, $S_0'$ is also a $k$-module of $H_B$. Observe that $S_0' \subseteq S' \subseteq S_0' \cup W_{r'+1}$. Finally, since we have $(W_{r'+1},c) \in {\cal L}_B$ {\em and} by the hypothesis, no $s_i$ has a least common ancestor with $c$ which is a strict descendant of $s$, we {\em cannot} have $W_{r'+1} \subseteq S'$. As a result, $S' = S_0'$.
   
\item Otherwise, let us assume w.l.o.g. that the least common ancestor of $c$ and $s_1$ is a strict descendant of $s$. Let us call it $t_1$. 

\begin{claim}\label{claim:b-2-1}
$t_1$ is a child of $s$ in $T^U$.
\end{claim}
\begin{proofclaim}
There are two sub-cases. First, let us assume $s_1 = t_1$. If $s_1$ is not a child of $s$ in $T^U$ then its former father node, call it $s_1'$, got removed when we created $T^B$. Then, either $s_1' = c$, or $s_1'$ is the father of $c$ in $T^U$. In both cases, this contradicts our assumption that $s_1$ is an ancestor of $c$ in $T$. 
Thus, from now on, let us assume $t_1 \neq s_1$. Since the father of $s_1$ in $T^B$ is $s$, $t_1$ got removed at some earlier recursive stage. In fact, this must be when we created $T^B$ because we have $s_1,c \in V(T^U)$ (otherwise, if it were done earlier, we could have not removed $t_1$ since it still had at least two children). Then again, either $t_1 = c$, or $t_1$ is the father of $c$ in $T^U$. Suppose for the sake of contradiction that $s$ is not the father of $t_1$ in $T^U$. Then, at least two nodes got removed from $T^U_c \setminus \left(\bigcup_{i=1}^p T^U_{a_i} \right)$ in order to create $T^B$. This can happen only if $c$ became a leaf, and then the two nodes removed must be $c$ and its father in $T^U$. But then, we should have $t_1 = c$, that contradicts the fact that $c$ became a leaf. 
\end{proofclaim}

We can also prove, as another intermediate claim (similar to the above Claim~\ref{claim:b-2-1}), that every node $s_i, \ i > 1$, is a child of $s$ in $T^U$. Indeed, if it were not the case for some $s_i$ then its father node $t_i$ in $T^U$ got removed when we created $T^B$. We either have $t_i = c$ or $t_i$ is the father of $c$ in $T^U$. In particular, $t_1$ is an ancestor of $t_i$. However, since $s_1,s_i \in V(T^B)$, this would contradict the removal of $t_1$ from $T^B$.
Overall, we proved as claimed that $t_1$ and $s_2,s_3,\ldots,s_q$ are children of $s$ in $T^U$. In particular, $f^U(t_1) = A \cup S_1$, while for every $2 \leq i \leq q$, $f^U(s_i) = f^B(s_i) = S_i$. 
Let $S_0'$ be the union of $A \cup S$ with all the clusters $W_j$, for $(W_j,c_j) \in {\cal L}$, s.t. the least common ancestor in $T$ of $c_j$ and some node $s_i$ is a strict descendant of $s$. By the induction hypothesis, $S_0'$ is a $k$-module of $H$. Furthermore, 

\begin{claim}\label{claim:b-2-2}
{\em Every} $(W_j,c_j) \in {\cal L} \cap {\cal L}_A$ satisfies $W_j \subseteq S_0'$.
\end{claim}
\begin{proofclaim}
If $(W_j,c_j) \in {\cal L} \cap {\cal L}_A$, then there exists an $1 \leq i' \leq p$ s.t. the least common ancestor of $c_j$ and $a_{i'}$ is a strict descendant of $c$. In particular, the least common ancestor of $s_1$ and $c_j$ is a strict descendant of $s$.
\end{proofclaim}

We get by Claim~\ref{claim:b-2-2} that $A' \subseteq S_0'$. Let $B' = S_0' \setminus A'$. Since $A' \cup B'$ is a $k$-module of $H$, by Lemma~\ref{lem:cut-preserver}, $W_{r'+1} \cup B'$ is a $k$-module of $H_B$. Finally, we observe that $S' = W_{r'+1} \cup B'$. 
\end{itemize}

\paragraph{Complexity analysis.} 
By induction, for every $r \geq 0$, for every $\langle r, H, U, T^U, f^U, {\cal L} \rangle$, we have $|U| \leq (2/3)^rn$. In particular, the depth of the recursion tree is ${\cal O}(\log{n})$ (as it was anticipated when we presented above the algorithm). Furthermore, for any fixed $r$, if we consider the sets $U$ of all the inputs $\langle r, H, U, T^U, f^U, {\cal L} \rangle$, then we get a (possibly partial) partition of $V$. In particular, the sum of all the values  $n_r = |V(H)|$, over all the inputs $\langle r, H, U, T^U, f^U, {\cal L} \rangle$ that are at the same recursion level $r$, is at most $n + n \times {\cal O}(rk^2) = {\cal O}(k^2n\log{n})$. In the same way, the sum of all the values  $m_r = |E(H)|$, over all the inputs $\langle r, H, U, T^U, f^U, {\cal L} \rangle$ that are at the same recursion level $r$, is at most $m + n \times {\cal O}(k^2r) = {\cal O}(k^2n\log{n} + m)$.  Processing $\langle r, H, U, T^U, f^U, {\cal L} \rangle$ takes ${\cal O}(2^{{\cal O}(k)}(n_r+m_r)^{1+o(1)})$ time if we exclude the recursive calls. Therefore, the total running time at any fixed recursive stage, and so also for the whole algorithm, is in ${\cal O}(2^{{\cal O}(k)}(n+m)^{1+o(1)})$.
\end{proof}

\section{Open problem}\label{sec:pb}

Our main contribution in this work is a quasi linear-time parameterized algorithm for computing the diameter of a graph, with singly-exponential dependency on the clique-width. This is optimal assuming SETH. For the parameter modular-width, for short {\tt mw}, there is an ${\cal O}(|V| + |E| + {\tt mw}^3)$-time algorithm in order to compute all the eccentricities~\cite{CDP19}. {\em Shrub-depth} is sometimes regarded as an interesting competitor for modular-width~\cite{gajarsky2013parameterized}. We observe that its algorithmic applications to polynomial-time solvable problems have yet to be explored. In particular, given a $(d,m)$-tree model for a graph $G=(V,E)$, can we compute $diam(G)$ in ${\cal O}(poly(d,m) \cdot (|V| + |E|)^{2-\epsilon})$ time, for some $\epsilon > 0$?   

\bibliographystyle{abbrv}
\bibliography{cw}

\end{document}